\documentclass[11pt]{article}

\usepackage{graphicx} % standard LaTeX graphics tool
\usepackage{multicol} % used for the two-column index
\usepackage{cropmark} % cropmarks for pages without
\usepackage{amssymb,amsmath,amscd}
\input epsf.tex

\newcommand{\CR}{\hbox{{$\cal R$}}} % curly font
\newcommand{\CF}{\hbox{{$\cal F$}}}
\newcommand{\CH}{\hbox{{$\cal H$}}}
\newcommand{\CG}{\hbox{{$\cal G$}}}
\newcommand{\CC}{\hbox{{$\cal C$}}}
\newcommand{\CA}{\hbox{{$\cal A$}}}
\newcommand{\cm}{\mathfrak{m}}  % lie algebras
\newcommand{\cg}{\mathfrak{g}}
\newcommand{\cb}{\mathfrak{b}}
\newcommand{\Q}{\mathbb{Q}}  % open font
\newcommand{\R}{\mathbb{R}}
\newcommand{\C}{\mathbb{C}}

\newcommand{\h}{{\scriptstyle\frac{1}{2}}}
\newcommand{\extd}{{\rm d}}
\newcommand{\del}{\partial}
\newcommand{\isom}{{\cong}}
\newcommand{\eps}{{\epsilon}}
\newcommand{\tens}{\mathop{\otimes}}
\newcommand{\la}{{\triangleright}}

\newcommand{\grav}{\mathsf{G}}
\newcommand{\Ext}{{\rm Ext}}
\newcommand{\id}{{\rm id}}
\newcommand{\<}{\langle}
\renewcommand{\>}{\rangle}

\newcommand{\link}{{\rm link}}

\newcommand{\und}[1]{{\underline {#1}}}
\newcommand{\eqn}[2]{\begin{equation}#2\label{#1}\end{equation}}

\newcommand{\rcross}{{\triangleright\!\!\!<}}
\newcommand{\lcross}{{>\!\!\!\triangleleft}}
\newcommand{\rcocross}{{\blacktriangleright\!\!<}}

\newcommand{\cobicross}{{\triangleright\!\!\!\blacktriangleleft}}
\newcommand{\bicross}{{\blacktriangleright\!\!\!\triangleleft}}
\newcommand{\dcross}{{\bowtie}}
\newcommand{\lbiprod}{{>\!\!\!\triangleleft\kern-.33em\cdot}}
\def\rbiprod{{\cdot\kern-.33em\triangleright\!\!\!<}}

\newtheorem{proposition}{Proposition}
\newtheorem{example}{Example}
\newtheorem{theorem}{Theorem}
\newtheorem{conjecture}{Conjecture}

\begin{document}\baselineskip 17pt

{\ }\qquad \hskip 4.3in
\vspace{.2in}

\begin{center} {\LARGE Meaning of Noncommutative Geometry and the Planck-Scale Quantum Group}
\\ \baselineskip 13pt{\ }\\
{\ }\\ Shahn Majid \\ {\ }\\ Department of Applied Mathematics and Theoretical
Physics\\ University of Cambridge, Silver Street, CB3 9EW, UK
\end{center}
\begin{center}
July, 1999
\end{center}
%\vspace{10pt}

\begin{quote}\baselineskip 13pt
This is an introduction for nonspecialists to the noncommutative
geometric approach to Planck scale physics coming out of quantum
groups. The canonical role of the `Planck scale quantum group'
$\C[x]\bicross\C[p]$ and its observable-state T-duality-like
properties are explained. The general meaning of noncommutativity
of position space as potentially a new force in Nature is
explained as equivalent under quantum group Fourier transform to
curvature in momentum space. More general quantum groups
$\C(G^\star)\bicross U(\cg)$ and $U_q(\cg)$ are also discussed.
Finally, the generalisation from quantum groups to general quantum
Riemannian geometry is outlined. The semiclassical limit of the
latter is a theory with generalised non-symmetric metric
$g_{\mu\nu}$ obeying $\nabla_\mu g_{\nu\rho}-\nabla_\nu
g_{\mu\rho}=0$.
\end{quote}

\section{Introduction} \baselineskip 17pt

There are currently several approaches to Planck-scale physics and
of them `Noncommutative geometry' is probably the most radical but also the
least well-tested. As Lee Smolin in his lectures at the conference
was kind enough to put it, it is `promising
but too early to tell'. Actually my point of view, which I will
explain in these lectures, is that some kind of noncommutative
geometry is {\em inevitable} whatever route we take to the Planck scale.
Whether we evolve our understanding of string theory, compute
quasiclassical states in loop-variable quantum gravity, or just
investigate the intrinsic mathematical structure of geometry and
quantum theory themselves (my own line), all roads will in my
opinion lead to {\em some kind} of noncommutative geometry as the
next more general geometry beyond nonEuclidean that is needed at
the Planck scale where both quantum and gravitational effects are
strong. I think the need for this and its general features can be
demonstrated from simple nontechnical arguments and will try to do
this here. These philosophical and conceptual issues are in
Section~2.

Beyond this, and definitely a matter of opinion, it seems to me
that there is are certain philosophical principles \cite{Ma:pri}
which can serve as a guide to what Planck scale physics should be,
in particular what I have called the {\em principle of
representation-theoretic self-duality} (of which T-duality is one
manifestation). I believe that to proceed one has to ask in fact
what is the {\em nature of physical reality itself}. In fact I do
not think that theoretical physicists can any longer afford to shy
away from such questions and, indeed, with proposals for
Planck-scale physics beginning to emerge it is already clear that
some new philosophical basis is going to be needed which will
likely be every bit as radical as those that came with quantum
mechanics and general relativity. My own radical philosophy in
\cite{Ma:the}\cite{Ma:pla}\cite{Ma:pri} basically takes the
reciprocity ideas of Mach to a modern setting. But it also
suggests a different concept of reality, which I call {\em relative
realism}, from the reductionist one that most theoretical
physicists are still unwilling to give up (I said it would be
radical). This might seem fanciful but what it boils down to
in practice is an extension of ideas of Fourier theory to
the quantum domain. Section~3 provides a modern introduction to this.

Next I will try to convince you that while there are still several
different ideas for what {\em exactly} noncommutative geometry
should be, there is slowly emerging what I call the `quantum
groups approach to noncommutative geometry' which is already {\em
fairly} complete in the sense that it has the same degree of
`flabbiness' as Riemannian geometry (is not tied to specific
integrable systems etc.) while at the same time it includes the
`zoo' of already known naturally occurring examples, mostly linked
to quantum groups. Picture yourself for a moment in the times of
Gauss and Riemann; clearly spheres, tori, etc., were evidently
examples of something, but of what? In searching for this Riemann
was able to formulate the notion of Riemannian manifold as a way
to capture known examples like spheres and tori but broad enough
to formulate general equations for the intrinsic structure of
space itself (or after Einstein, space-time). Theoretical physics
today is in a parallel situation with many naturally occurring
examples from a variety of sources and a clear need for a general
theory. Our approach\cite{Ma:rie} is based on fiber bundles with
quantum group fiber\cite{BrzMa:gau}, and we will come to it by the
end of the lectures, in Section~6. It includes a working
definition of `quantum manifold'.

In between, I will try to give you a sense of some noncommutative
geometries out there from which our intuition has to be drawn. We
will `see the sights' in the land of noncommutative geometry at
least from the quantum groups point of view. Just as Lie groups
are the simplest Riemannian manifolds, quantum groups are the
simplest noncommutative spaces. Their homogeneous spaces are also
covered, as well as quantum planes (which are more properly
braided groups). We refer to \cite{Ma:book} for more on quantum groups
themselves.

At the same time, the physics reader will no doubt also want to
see testable predictions, detailed models etc. While, in my
opinion, it is {\em still} too early to rush into building models and
making predictions (`one cannot run before one can walk') I will
focus on at least one toy model of Planck-scale physics using
these techniques. This is the Planck-scale quantum group
introduced 10 years ago in \cite{Ma:pla}\cite{Ma:the} and
exhibiting even then many of the features one might consider
important for Planck scale physics today, including duality. This is the
topic of Section~4. It is not, however, the `theory of everything'
or M-theory etc. I seriously doubt that Einstein could have
formulated general relativity without the mathematical definition
of a `manifold' having been sorted out by Riemann a century before
(and which I would guess had filtered down to Einstein's mathematical
mentors such as Minkowski). In the same way, one really needs to
sort out the correct or `natural' definitions of noncommutative
geometry some more (in particular the Ricci tensor and stress
energy tensor are not yet understood) before making attempts at a
full theory with testable predictions. This is on the one hand
mathematics but on the other hand it has to be guided by physical
intuition with or even without firm predictive models. In fact the
structure of the mathematical possibilities of noncommutative
geometry (which means for us results in the theory of algebra) can
tell us a lot about any actual or effective theory even if it is
not presently known.

The general family of {\em bicrossproduct quantum groups} arising
in this way out of Planck scale physics contains many more
examples (it is one of the two main constructions by which quantum
groups originated in physics.) For example, there is a quantum
group $\C(G^\star)\bicross U(\cg)$ for every complex simple Lie
algebra $\cg$. All these
 bicrossproduct quantum groups can be viewed as the
actual quantum algebras of observables of actual quantum systems and can
be viewed precisely as models unifying quantum and gravity-like effects
\cite{Ma:the}\cite{Ma:pla}. For the record, the bicrossproduct construction
$\bicross$ was introduced in this context at about the same time (in 1986)
but independently of the more well-known quantum groups $U_q(\cg)$
\cite{Dri}\cite{Jim:dif}, in particular before I had even heard of V.G. Drinfeld
or integrable systems. To this day the two classes of quantum groups, although
constructed from the same data $\cg$, have never been directly related (this remains
an interesting open problem). The situation is shown in Figure~1. To build
a theory of noncommutative geometry we need to include naturally occurring examples
such as these.

We also need to include the more traditional noncommutative algebras
to which people have traditionally tried to develop geometric pictures, namely
the canonical commutation relations algebra $[x,p]=\imath\hbar$ or its group version the
Weyl algebra or `noncommutative
torus' $ vu=e^{\imath\alpha}uv$ as in the work of A. Connes\cite{Con:geo}. We can also
consider the matrix algebras
$M_n(\C)$ as studied by Dubois-Violette, Madore and others; as we saw
seen in the beautiful lecture of Richard Kerner at the conference, one can do a certain amount
of noncommutative geometry for such algebras too. On the other hand, in
some sense these are actually all the same
example in one form or another, i.e. basically the algebra of operators on some
Hilbert space (at least for generic $\alpha$). These examples and the traditional
ideas of vector fields as derivations and points as maximal ideals etc., come
from algebraic geometry and predate
quantum groups. In my opinion, however, one cannot build a valid noncommutative
geometry always on the basis of essentially one example (and a lot of elegant
mathematics) -- one has to also include the
rich vein of practical examples such as the quantum groups above. The latter have a much clearer geometric
meaning but very few derivations or maximal ideals etc., i.e. we have to develop
a much less obvious noncommutative differential geometry if we are to include them as well
as the traditional matrix algebras and of course the commutative case corresponding to
usual geometry. This is precisely what has emerged slowly in recent years and that which I will try to explain.
\begin{figure}
\[ \epsfbox{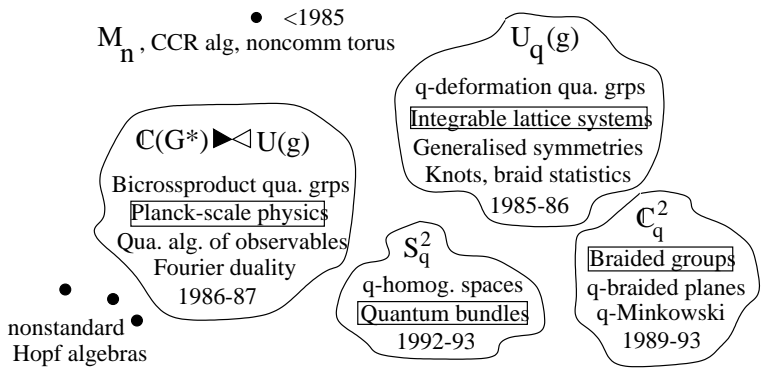}\]
\caption{The landscape of noncommutative geometry today. Some isolated `traditional'
objects such as matrix algebras and the noncomm. torus, and two classes of quantum
groups}
\end{figure}

In Section~5 we turn for completeness, to the other and more
well-known type of quantum groups, the q-deformed
enveloping algebras $U_q(\cg)$. These did not arise at all
in connection with Planck scale physics or even directly as the
quantisation of any physical system. Rather they are `generalised
symmetries' of quantum or lattice integrable systems. Nevertheless
they are also examples of noncommutative geometry and, if recent
conjectures of Lee Smolin and collaborators prove correct, 
they are natural
descriptions of the noncommutative geometry coming out of the loop
variable approach to quantum gravity. The more established meaning of
{\em these} quantum groups
which we will focus on is that they induce braid statistics on
particles transforming as their representations. In effect the
dichotomy of particles into bosons (force) and fermions (matter)
is broken in noncommutative geometry and in fact both are unified
with each other and with quantisation. Roughly speaking the meaning
of $q$ here is a
generalisation of the $-1$ for supersymmetry. So the natural
noncommutative geometry here is `braided
geometry'\cite{Ma:introp}. Yet at the same time one may write $q$
in terms of Planck's constant or, according to \cite{MajSmo:def}, the
cosmological constant. It means that one physical manifestation of
quantum gravity effects is as braid (e.g. fractional) statistics.

Finally, more accessible perhaps to many readers will be not so
much our proposals for the full noncommutative theory but its
semiclassical predictions; in order to be naturally made
noncommutative one has to shift ones point of view a little and
indeed move to a slightly more general notion of classical
Riemannian geometry. The main prediction is that one should
replace the notion of metric and its Levi-cevita connection by a
notion of nondegenerate 2-tensor (not necessarily symmetric) and
the notion of vanishing torsion and vanishing {\em cotorsion}. The
cotorsion tensor associated to a 2-tensor is a new concept recently
introduced in \cite{Ma:rie}. The resulting self-dual generalisation of the
usual metric compatibility becomes
\[ \nabla_\mu g_{\nu\rho}-\nabla_\nu g_{\mu\rho}=0.\]
The generalisation allows a synthesis of symplectic and Riemannian geometry,
which is a semiclassical analogue of the quantum-gravity unification problem.
Not surprisingly, the above ideas  turn out to be related at the
semiclassical level to other ideas for Planck scale physics such as
 T-duality for sigma models on Poisson-Lie groups\cite{Kli:poi}, 
see \cite{BegMa:poi}.

\section{The meaning of noncommutative geometry}

It stands to reason that if one seriously wants to unify quantum
theory and gravity into a single theory with a single elegant
point of view, one must first formulate each in the same language.
On the side of gravity this is perfectly well-known and we do not
need to belabour it; instead of points in a manifold one should
and does speak in terms of the algebra of its coordinate functions
e.g. (locally) position coordinates $\bf x$, say. Geometrical
operations can then be expressed in terms of this algebra, for
example a vector field might be a derivation on the algebra.
`Points' might be maximal ideals. This conventional point of view
(called {\em algebraic geometry}) doesn't really work in practice
in the noncommutative case, i.e. it needs to be modified, but it
is a suitable starting point for the unification.

What about quantum mechanics? Well this too is some kind of
algebra, of course noncommutative due to noncommutation relations
between position and momentum. So the language we need is that of
algebras. We need to modify usual algebraic geometry in such a way
that it extends to algebras of observables arising in quantum
systems. At the same time we should, I believe, also be guided by
finding natural mathematical definitions that both include
nontrivial applications in mathematics {\em and} encode those algebras
in quantum systems which have a clear geometrical structure
self-evidently in need of being encoded (perhaps even without direct
physical input about Planck scale physics). For example, before the
discovery of quantum groups noncommutative geometry made only
minimal changes in pursuit of the above idea, e.g. to let the
algebra be noncommutative but nevertheless define a vector field as a
derivation. All very elegant, but not sufficient to include `real
world' examples like quantum groups.

One other general point. For classical systems we frequently make use
of deep classification and other theorems about smooth manifolds; the
rich structure of what is mathematically allowed e.g. by topological
constraints is often a guide to building effective theories even
if we do not know the details of the underlying theory. If we accept
the above then the
corresponding statement is that deep mathematical theorems about
the classification and structure of noncommutative algebras ought
to tell us about the possible effective corrections from quantum
gravity even before a full theory is known (as well as be a guide
to the natural structure of the full theory). We will see this in some toy
examples in the next chapter. By contrast many physicists seem to
believe that the only algebra in physics is the CCR algebra (or
its fermionic version), or possibly Lie algebras, as if there is
not in fact a much richer world of noncommutative algebras for
their theories to draw upon. In fact this noncommutative world has
to be at least as rich as the theory of manifolds since it must
contain them in the special commutative limit. I contend that the
intrinsic properties of noncommutative algebras is where we should
look for new principles and ideas for the Planck scale.

\subsection{Curvature in momentum space -- a possible new force of nature}

Before going into details
 of the modern approach to noncommutative
geometry we want to consider some general issues about unifying
quantum theory and geometry using algebra. In particular, what
finally emerges as the true meaning of noncommutative geometry for
Planck scale physics? In a nutshell, the answer I believe is as
follows. Thus, to survive to the Planck scale we should cling to
only the very deepest ideas about the nature of physics. In my
opinion among the deepest is `Born reciprocity' or the arbitrariness
under position and momentum. Now, in conventional flat space quantum
mechanics we take the $\bf x$ commuting among themselves and their
momentum $\bf p$ likewise commuting among themselves. The
commutation relation
\eqn{heis}{ [x_i,p_j]=\imath\hbar\delta_{ij}}
is likewise symmetric in the roles of ${\bf x},{\bf p}$ (up to a sign).
To this symmetry
may be attributed such things as wave-particle duality. A wave
has localised $\bf p$ and a particle has localised $\bf x$.

Now the meaning of {\em curvature} in position space is, roughly
speaking, to make the natural conserved $\bf p$ coordinates
noncommutative. For example, when the position space is a 3-sphere
the natural momentum is $su_2$. The enveloping algebra $U(su_2)$
should be there in the quantum algebra of observables with relations
\eqn{su2p}{ [p_i,p_j]=\frac{\imath}{ R}\eps_{ijk}p_k}
where $R$ is proportional to the radius of curvature of the $S^3$.

By Born-reciprocity then there should be another possibility which
is {\em curvature in momentum space}. It corresponds under Fourier
theory to noncommutativity of position space. For example if the
momentum space were a sphere with $m$ proportional to the
radius of curvature, the position space
coordinates would correspondingly have noncommutation relations
\eqn{su2x}{ [x_i,x_j]=\frac{\imath}{ m}\eps_{ijk}x_k.}
Mathematically speaking this is surely a symmetrical and equally
interesting possibility which might have observable consequences
and might be observed. Note that $m$ here is just a parameter not
necessarily mass, but our use of it here does suggest the
possibility of understanding the {\em geometry of the mass-shell}
as noncommutative geometry of the position space $\bf x$. This may
indeed be an interesting and as yet unexplored application of
these ideas. In general terms, however, the situation is clear:
for systems constrained in position space one has the usual tools
of differential geometry, curvature etc., of the constrained
`surface' in position space or tools for noncommutative algebras
(such as Lie algebras) in momentum space. {\em For systems
constrained in momentum space one needs conventional tools of
geometry in momentum space or, by Fourier theory, suitable tools
of noncommutative geometry in position space.}

In mathematical terms, these latter two examples
(\ref{su2p}),(\ref{su2x}) demonstrate the point of view of
noncommutative geometry: we are viewing the enveloping algebra
$U(su_2)$ {\em as if} it were the algebra of coordinates of some
system, i.e. we want to answer the question
\[ U(su_2)=C(?)\]
where $?$ will not be any usual kind of space (where the
coordinates would commute). This is what we have called in
\cite{Ma:ista} a `quantum-geometry transformation' since a quantum
symmetry point of view (such as the angular momentum in a quantum
system) is viewed `up-side-down' as a geometrical one. The
simplest example $U(\cb_+)$ was studied from this point of view as
a noncommutative space in \cite{Ma:reg}, actually slightly more
generally as $U_q(\cb_+)$.

For particular examples of this type we do not of course
{\em need} any fancy noncommutative geometric point of view -- Lie
theory was already extensively developed just to handle such
algebras. But if we wish to unify quantum and geometric effects
then we should  start taking this noncommutative geometric
viewpoint even on such familiar algebras. What are `vector fields' on
$U(su_2)$? What is Fourier transform
\[ \CF:U(su_2)\to \C(SU_2)\]
from the momentum coordinates to the $SU_2$ position coordinates?
These are nontrivial (but essentially solved) questions.
Understanding them, we can proceed to construct more complex
examples of noncommutative geometry which are neither $U(\cg)$ nor
$C(G)$, i.e. where noncommutative geometry is really needed and
where both quantum and geometrical effects are unified. Vector
fields, Fourier theory etc., extend to this domain and allow us to
explore consistently new ideas for Planck scale physics. This
approach to Planck scale physics based particularly on Fourier
theory to extend the familiar $\bf x,p$ reciprocity to the case of
nonAbelian Lie algebras and beyond is due to the author in
\cite{Ma:the}\cite{Ma:phy}\cite{Ma:pla}\cite{Ma:pri}
\cite{Ma:ista} and elsewhere.

Notice also that the three effects exemplified by the three
equations (\ref{heis})--(\ref{su2x}) are all independent. They are
controlled by three different parameters $\hbar,R,m$ (say). Of course
in a full theory of quantum gravity all three effects could exist
together and be unified into a single noncommutative algebra generated by
suitable $\bf x,p$. Moreover, even if we do not know the details of the correct
theory of quantum gravity, if we assume that something like Born
reciprocity survives then all three effects indeed {\em should}
show up in the effective theory where we consider almost-particle
states with position and momenta $\bf x,\bf p$. It would require
fine tuning or some special principle to eliminate any one of them.
Also the same ideas could apply at the level of the quantum
gravity field theory itself, but this is a different
question.

\subsection{Algebraic structure of quantum mechanics}

In the above discussion we have assumed that quantum systems are
described by algebras generated by position and momentum. Here we
will examine this a little more closely. The physical question to
keep in mind is the following: {\em what happens to the geometry
of the classical system when you quantise?}

To see the problem consider what you obtain when you quantise a
sphere or a torus. In usual quantum mechanics one takes the
Hilbert space on position space, e.g. $\CH=L^2(S^2)$ or
$\CH=L^2(T^2)$ and as `algebra of observables' one takes
$A=B(\CH)$ the algebra of all bounded (say) operators. It is
decreed that every self-adjoint hermitian operator $a$ (or its
bounded exponential more precisely) is an observable of the system
and its expectation value in state $|\psi>\in \CH$ is
\[ <a>_\psi=<\psi|a|\psi>.\]
The problem with this is that $B(\CH)$ {\em is the same algebra in
all cases}. The quantum system does know about the underlying
geometry of the configuration space or of the phase space in other
ways; the choice of `polarisation' on the phase space or the
choice of Hamiltonian etc. -- such things are generally defined
using the underlying position or phase space geometry -- but the
abstract algebra $B(\CH)$ doesn't know about this. All separable
Hilbert spaces are isomorphic (although not in any natural way) so
their algebras of operators are also all isomorphic. In other
words, whereas in classical mechanics we use extensively the
detailed geometrical structure, such as the choice of phase space
as a symplectic manifold, all of this is not recorded very
directly in the quantum system. One more or less forgets it,
although it resurfaces in relation to the more restricted kinds of
questions (labeled by classical `handles') one asks in practice
about the quantum system. In other words, {\em the true quantum
algebra of observables should not be the entire algebra $B(\CH)$
but some subalgebra $A\subset B(\CH)$}. The choice of this
subalgebra is called the {\em kinematic structure} and it is
precisely here that the (noncommutative) geometry of the classical
and quantum system is encoded. This is somewhat analogous to the
idea in geometry that every manifold can be visualised concretely
embedded in some $\R^n$. Not knowing this and thinking that
coordinates $\bf x$ were always globally defined would miss out on
all physical effects that depend on topological sectors, such as
the difference between spheres and tori.

Another way to put this is that by the Darboux theorem all
symplectic manifolds are {\em locally} of the canonical form
$\extd x\wedge \extd p$ for each coordinate. Similarly one should
take (\ref{heis}) (which essentially generates all of $B(\CH)$,
one way or another) only locally. The full geometry in the quantum
system is visible only by considering more nontrivial algebras
than this one to bring out the global structure. We should in fact
consider all noncommutative algebras equipped with certain
structures common to all quantum systems, i.e. inspired by
$B(\CH)$ as some kind of local model or canonical example but not
limited to it. The conditions on our algebras should also be
enough to ensure that there {\em is} a Hilbert space around and
that $A$ can be viewed concretely as a subalgebra of operators on
it.

Such a slight generalisation of quantum mechanics which allows
this kinematic structure to be exhibited exists and is quite
standard in mathematical physics circles. The required
algebra is a {\em von Neumann} algebra or, for a slightly nicer
theory, a $C^*$-algebra. This is an algebra over $\C$ with a $*$
operation and a norm $||\ ||$ with certain completeness and other
properties. The canonical example is $B(\CH)$ with the operator
norm and $*$ the adjoint operation, and every other is a
subalgebra.

Does this slight generalisation have observable consequences?
Certainly. For example in quantum statistical mechanics one
considers not only state vectors $|\psi>$ but `density matrices'
or generalised states. These are convex linear combinations of the
projection matrices or expectations associated to state vectors
$|\psi_i>$ with weights $s_i\ge 0$ and $\sum_i s_i=1$. The
expectation value in such a `mixed state' is
\eqn{dens}{ <a>=\sum_i s_i<\psi_i|a|\psi_i>}
In general these possibly-mixed states are equivalent to simply specifying
the expectation directly as a linear map $<\
>:B(\CH)\to \C$. This map respects the adjoint or $*$ operation on
$B(\CH)$ so that $<a^*a>\ge 0$ for all operators $a$ (i.e. a positive linear functional)
and is also continuous with respect to the operator norm. Such
positive linear functionals on $B(\CH)$ are precisely of the above
form (\ref{dens}) given by a density matrix, so this is a complete
characterisation of mixed states with reference only to the
algebra $B(\CH)$, its $*$ operation and its norm. The expectations
$<\ >_\psi$ associated to ordinary Hilbert space states are called the
`pure states' and are recovered as the extreme points in the
topological space of positive linear functionals (i.e. those which
are not the convex linear combinations of any others).

Now, if the actual algebra of observables is some subalgebra
$A\subset B(\CH)$ then any positive linear functional on the
latter of course restricts to one on $A$, i.e. defines an
`expectation state' $A\to \C$ which associates numbers, the
expectation values, to each observable $a\in A$. But not
vice-versa, i.e. the algebra $A$ may have perfectly well-defined
expectation states in this sense which are not extendable to all
of $B(\CH)$ in the form (\ref{dens}) of a density matrix.
Conversely, a pure state on $B(\CH)$ given by $|\psi>\in\CH$ might
be mixed when restricted to $A$. The distinction becomes crucially
important for the correct analysis of quantum thermodynamic
systems for example, see \cite{BraRob:ope}.

The analogy with classical geometry is that not every local
construction may be globally defined. If one did not understand
that one would miss such important things as the Bohm-Aharanov
effect, for example. Although I am not an expert on the `measurement problem'
in the philosophy of quantum mechanics it does not surprise me that
one would get into inconsistencies if one did not realise that the
algebra of observables is a subalgebra of $B(\CH)$. And from our
point of view it is precisely to understand and `picture' the
structure of the subalgebra for a given system that noncommutative
geometry steps in. I would also like to add that the problem of
measurement itself is a matter of matching the quantum system to
macroscopic features such as the position of measuring devices. I
would contend that to do this consistently one first has to know
how to identify aspects of `macrospopic structure' in the quantum
system without already taking the classical limit. Only in this
way can one meaningfully discuss concepts such as partial
measurement or the arbitrariness of the division into measurer
and measured. Such an
identification is exactly the task of noncommutative geometry,
which deals with extending our macroscopic intuitions and
classical `handles' over to the quantum system. Put another way,
the correspondence principle in quantum mechanics typically involves
choosing local coordinates like $\bf x,p$ to map over. Its refinement to
correspond more of the global geometry into the quantum world is
the practical task of noncommutative geometry.

\subsection{Principle of
representation-theoretic self-duality}

With the above preambles, we are in a position to consider some
speculation about Planck scale physics. Personally I believe that
anything we write down that is based on our past experience and
not on the deepest philosophical principles is not likely to
survive except as an approximation. For example, while string
theory may indeed survive to models of the Planck scale as a
certain approximation valid in a certain domain, it does not have
enough of a radical new philosophy to provide the true conceptual
leaps. I should apologise for this belief but I do not believe
that Nature cares about the historically convenient route by which
we might arrive at the right concepts for the Planck scale.

So as a basis we should stick only to some of the deepest principles. In my opinion
one of the deepest principles concerns the nature of mathematics itself. Namely
throughout mathematics one finds an intrinsic dualism between observer and observed
as follows. When we think of a function $f$ being evaluated on $x\in X$, we
could equally-well think of the same numbers as $x$ being evaluated on $f$ a member
of some dual structure $f\in\hat X$:
\[ {\rm Result}=f(x)=x(f).\]
Such a `turning of the tables' is a mathematical fact. For any
mathematical concept $X$ one may consider maps or
`representations' from it to some self-evident class of objects
(say rational numbers or for convenience real or complex numbers)
wherein our results of measurements are deemed to lie. Such
representations themselves form a dual structure $\hat X$ of
which elements of $X$ can be equally well viewed as representations. {\em But
is such a dual structure equally real?} I postulated in 1987
that indeed it should be so in a complete theory. Indeed\cite{Ma:pri}, {\em The
search for a complete theory of physics is the search for a
self-dual formulation in the above representation-theoretic sense
(The principle of representation-theoretic self-duality).} Put
another way, a complete theory of physics should admit a
`polarisation' into two halves each of which is the set of
representations of the other. This division should be arbitrary -- one
should be able to reverse interpretations (or indeed consider
canonical transformations to other choices of `polarisation' if
one takes the symplectic analogy).

Note that by completeness here I do not mean knowing in more and
more detail
{\em what} is true in the real world. That consists of greater and
greater complexity
but it is not {\em theoretical} physics. I'm considering that a theorist wants
to know {\em why} things are the way they are. Ideally I would like on my deathbed to
be able to say that I have found the right point of view or theoretical-conceptual
framework from which everything else follows. Working out the details of that would
be far from trivial of course. This is a more or less conventional reductionist viewpoint
except that the Principle asserts that we will not have found the required
 point of view unless
it is self-dual.

For example, there is a sense in which geometry -- or `gravity' is dual to
quantum theory or matter. This is visible for some simple models such as
spheres with constant curvature where it is achieved by Fourier theory. We will
be saying more about this later. If we accept this then in general terms Planck
scale physics has to unify these mutually dual concepts into one self-dual
structure. Ideally then Einstein's equation
\eqn{einst}{ G_{\mu\nu}=T_{\mu\nu}}
would appear as some kind of {\em self-duality} equation within this self-dual
context. Here the stress-energy tensor $T_{\mu\nu}$ measures how matter responds
to the geometry, while the Einstein tensor $G_{\mu\nu}$ measures how geometry responds
to matter. This is the part of Mach's principle which apparently inspired Einstein.
The question is how to make these ideas precise in a representation-theoretic sense.
While this still remains a long-term goal or vision, there are some toy models\cite{Ma:pla}
where some of the required features can be seen. We come to them in a later section.
For the moment we note only that one needs clearly some kind of noncommutative geometry because
$T_{\mu\nu}$ should really be the quantum operator stress-energy and its coupling
to $G_{\mu\nu}$ through its expectation value is surely only the first approximation
or semiclassical limit of an operator version of (\ref{einst}). But an operator
version of $G_{\mu\nu}$ only makes sense in the context of noncommutative geometry.
What we would hope to find, in a suitable version of these ideas, is a self-dual
setting where there was a dual interpretation in which $T_{\mu\nu}$ was the Einstein
tensor of some dual system and $G_{\mu\nu}$ its stress-energy. In this way the
duality and self-duality of the situation would be made manifest.
\begin{figure}
\[ \epsfbox{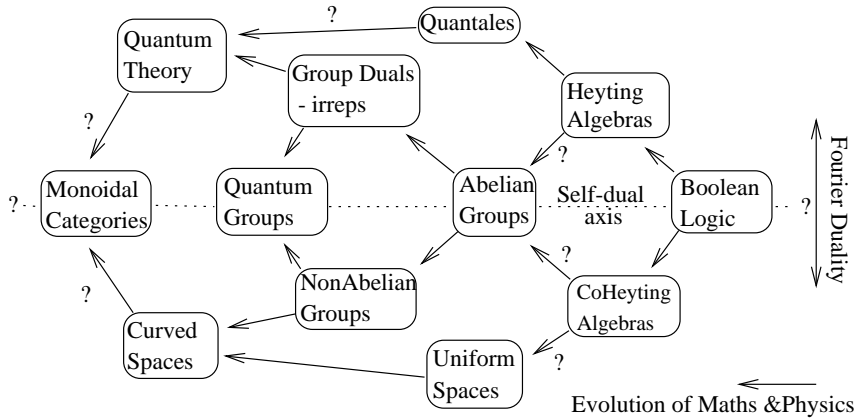}\]
\caption{Representation-theoretic approach to Planck-scale physics. The unification
of quantum and geometrical effects is a drive to the self-dual axis. Arrows
denote inclusion functors}
\end{figure}

This is more or less where quantum groups come in, as a simple and
soluble version of the more general unification problem. The
situation is shown in Figure~2. Thus, the simplest theories of
physics are based on Boolean algebras (a theory consists of
classification of a `universe' set into subsets); there is a
well-known duality operation interchanging a subset and its
complement. The next more advanced self-dual category is that of
(locally compact) Abelian groups such as $\R^n$. In this case the
set of 1-dimensional (ir)reps is again an Abelian group, i.e. the
category of such objects is self-dual. In the topological setting
one has $\hat\R^n\isom \R^n$ so that these groups (which are at
the core of linear algebra) are self-dual objects in the self-dual
category of Abelian groups. Of course, Fourier theory interchanges
these two. More generally, to accommodate other phenomena we step
away from the self-dual axis. Thus, nonAbelian Lie groups such as
$SU_2$ as manifolds provide the simplest examples of curved
spaces. Their duals, which means constructing irreps, appear as
central structures in quantum field theory (as judged by any
course on particle physics in the 1960's). Wigner even defined a
particle as an irrep of the Poincar\'e group. The unification of
these two concepts, groups and groups duals was for many years an
open problem in mathematics. Hopf algebras or quantum groups had
been invented as the next more general self-dual category
containing groups and group duals (and with Hopf algebra duality
reducing to Fourier duality) back in 1947 but no general classes
of quantum groups going beyond groups or group duals i.e. truly
unifying the two were known. In 1986 it was possible to view this
open problem as a `toy model' or microcosm of the problem of
unifying quantum theory and gravity and the bicrossproduct quantum
groups such as $\C(G^\star)\bicross U(\cg)$ were introduced on
this basis as toy models of Planck scale physics\cite{Ma:pla}. The
construction is self-dual (the dual is of the same general form).
At about the same time, independently, some other quantum groups
$U_q(\cg)$ were being introduced from a different point of view
both mathematically and physically (namely as generalised
symmetries). We go into details in later sections.

We end this section with some promised philosophical remarks. First of all, why the
principle of self-duality? Why such a central role for
Fourier theory? The answer I believe is that something very general like this (see the
introductory discussion) underlies the very nature of what it means to do science.
My model (no doubt a very crude one but which I think captures some of the essence
of what is going one) is as follows. Suppose that some theorist puts forward a theory in which
there is an actual group $G$ say `in reality' (this is where physics differs from math)
and some experimentalists construct tests of the theory and in so doing they routinely
build representations or elements of $\hat G$. They will end up regarding $\hat G$ as
`real' and $G$ as merely an encoding of $\hat G$. The two points of view are in harmony
because mathematically (in a topological context)
\[ G\isom\hat{\hat{G}}.\]
So far so good, but through the interaction and confusion between the experimental
and theoretical points of view one will eventually have to consider both, i.e. $G\times
\hat G$ as real. But then the theorists will come along and say that they don't like
direct products, everything should interact with everything else, and will seek
to unify $G,\hat G$ into some more complicated irreducible structure $G_1$, say. Then
the experimentalists build $\hat G_1$ ... and so on. This is a kind
of engine for the evolution of Science.

For example, if one regarded, following Newton that space $\R^n$ is real, its
representations $\hat \R^n$ are derived quantities ${\bf p}=m\dot {\bf x}$.
But after making diverse such representations one eventually regards
both $\bf x$ and $\bf p$ as equally valid, equivalent via Fourier theory.
But then we seek to unify them and introduce the CCR algebra
(\ref{heis}). And so on. Note that this is not intended to be a historical
account but a theory for how things should have gone in an ideal
case without the twists and turns of human ignorance.

One could consider this point of view as window dressing. Surely
quantum mechanics was `out there' and would have been discovered
whatever route one took? Yes, but if if the mechanism is correct
even as a hindsight, the same mechanism {\em does have predictive
power} for the next more complicated theory. The structure of the
theory of self-dual structures is nontrivial and not everything is
possible. Knowing what is mathematically possible and combining
with some postulates such as the above is not empty. For example,
back in 1989 and motivated in the above manner it was shown that
the category of monoidal categories (i.e. categories equipped with
tensor products) was itself a monoidal category, i.e. that there
was a construction $\hat{\CC}$ for every such category
$\CC$\cite{Ma:rep}. Since then it has turned out that both
conformal field theory and certain other quantum field theories
can indeed be expressed in such categorical terms. Geometrical
constructions can also be expressed categorically\cite{Ma:som}. On
the other hand, this categorical approach is still under-developed
and its exact use and the exact nature of the required duality as
a unification of quantum theory and gravity is still open. I would
claim only `something like that' (one should not expect too much
from philosophy alone).

Another point to be made from Figure~2 is that if quantum theory and gravity already
take us to very general structures such as categories themselves for the unifying
concept then, in lay terms, what it means is that the required theory involves very
general concepts indeed of a similar level to semiotics and linguistics (speaking
about categories of categories etc.). It is almost impossible to conceive {\em within
existing mathematics} (since it is itself founded in categories) what fundamentally
 more general structures would come after that. In other words, {\em the
required  mathematics is running out}
it least in the manner that it was developed in this century (i.e. categorically)
and at least
in terms of the required higher levels of generality in which to look for
self-dual structures. If the search for the ultimate theory of physics is to be restricted to logic
and  mathematics
(which is surely what distinguishes science from, say, poetry), then this
indeed
correlates with our physical intuition that the unification of quantum theory
and gravity is the last big unification for physics as we know it, or as were
that theoretical physics as we know it is coming to an end. I would agree with this
assertion except to say that the new theory will probably open up more questions
which are currently considered metaphysics and make them physics, so I don't
really think we will be out of a job even as theorists (and there will always
be an infinite amount of `what' work to be done even if the `why' question
was answered at some consensual level).

As well as seeking the `end of physics', we can also ask more about its birth. Again
there are many nontrivial and nonempty questions raised by the self-duality postulate.
Certainly the key generalisation of Boolean logic to intuitionistic logic is
to relax the axiom that $a\cup\tilde a=1$ (that $a$ or not $a$ is true). Such an algebra
is called a {\em Heyting algebra} and can be regarded as the birth of quantum mechanics.
Dual to this is the notion of a {\em coHeyting algebra} in which we relax the law
that $a\cap\tilde a=0$. In such an algebra one can define the `boundary' of a proposition
as
\[ \del a=a\cap \tilde a\]
and show that it behaves like a derivation. This is surely the
birth of geometry. How exactly this complementation duality
extends to the Fourier duality for groups and on to the duality
between more complex geometries and quantum theory is not
completely understood, but there are conceptual `physical'
argument that this should be so, put forward in \cite{Ma:pri}.

Briefly, in the simplest `theories of physics' based only on logic
one can work equally well with `apples' or `not-apples' as the
names of subsets. {\em What happens to this complementation
duality in more advanced theories of physics?} Apples curve space
while not-apples do not, i.e. in physics one talks of apples as
really existing while not-apples are merely an abstract concept.
Clearly the self-duality is lost in a theory of gravity alone. But
we have argued\cite{Ma:pri} that when one considers both gravity
{\em and} quantum theory the self-duality can be restored. Thus when we
say that a region is as full of apples as General Relativity
allows (more matter simply forms a black hole which expands),
which is the right hand limiting line in
Figure~3\footnote{Diagrams similar to the right hand side of
Figure~3 plotting the mass-energy density v size have been
attributed to Brandon Carter (who was here at the conference), as
a tool to plot stellar evolution}, in the dual theory we might say
that the region is as empty of not-apples as quantum theory
allows, the limitation being the left slope in Figure~3. Here the
uncertainty principle in the form of pair creation ensures that
space cannot be totally empty of `particles'. Although heuristic,
these are arguments that quantum theory and gravity are dual and
that this duality is an extension of complementation duality. Only
a theory with both would be self-dual. Also, in view of a `hole'
moving in the opposite direction to a particle, the dual theory
should also involves time reversal. The self-duality is something
like CPT invariance but in a theory where gravitational and not
only quantum effects are considered. We are proposing it as a key
requirement for quantum-gravity.

\begin{figure}
\[ \epsfbox{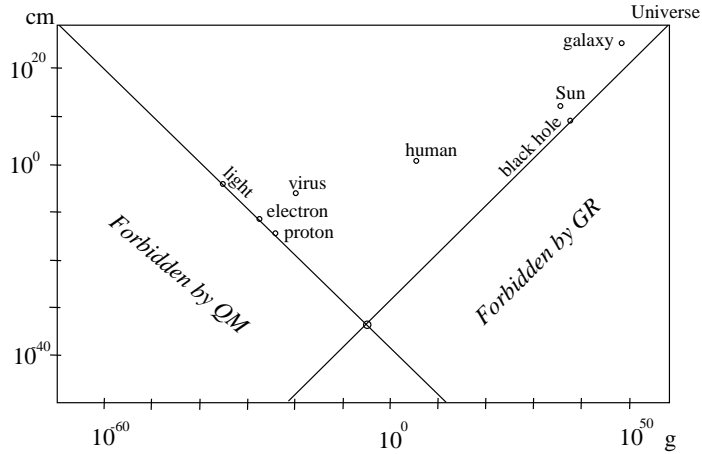}\]
\caption{Range of physical phenomena, which lie in the wedge
region with us in the middle. Log plots are mass-energy v size}
\end{figure}

\subsection{Relative realism}

So far we have given arguments that there is at least a
correlation between the mathematical structure of self-dual
structures and the progressive theories of physics from their
birth in `logic' to the projected forthcoming complete theory of
everything. It should at least provide a guide to the properties
that should be central in unknown theories of everything, such as
what have become fashionable to call `M-theory'.

What about going further? This section will indeed be speculative
but I believe it should be considered. Suppose indeed that some
mathematical-structural principles (such as the principle of
representation theoretic self-duality above) could exactly pin
down the ultimate theory of physics along the lines discussed.
This would be like giving a list of things that we expect from a
complete theory -- such as renormalisability, CPT-invariance, etc.,
except that we are considering such general versions of these
`constraints' that they are practically what it means to be a
group of people following the scientific method. If this really
pins down the ultimate theory then it means that {\em the ultimate
theory of physics is no more and no less than a self-discovery
 of the constraints in thinking that are
taken on when one decides to look at the world as a physicist.}

If this sounds cynical it is not meant to be; it is merely a
Kantian or Hegelian basis of physical reality as opposed the more
conventional reductionist one that most physicists take for
granted. It does not mean that physics is arbitrary or random any
more than the different possible manifolds `out there' are
arbitrary. The space of all possible manifolds up to equivalence
has a deep and rich structure and feels every bit as real to
anyone who studies it; but it is a mathematical reality `created'
when we accept the axioms of a manifold. So what we are saying is
that there is not such a fundamental difference between
mathematical reality and physical reality. The main difference is
that mathematicians are aware of the axioms while physicists tend
to discover them `backwards' by theorising from experience. I call
this point of view {\em relative realism}\cite{Ma:pri}. In it, we
experience reality through choices that we have forgotten about at
any given moment. If we become aware of the choice the reality it creates
is dissolved or `unconstructed'. On the other hand, the reader will
say that the possibility of the theory of manifolds -- that the
game of manifold-hunting could have been played in the first place
-- is itself a reality, not arbitrary. It is, but at a higher
level: it is a concrete fact in a more general theory of possible
axiom systems of this type. To give another example, the reality
of chess is created once we chose to play the game. If we are
aware that it is a game, that reality is dissolved, but the rules
of chess remain a reality although not within chess but in the space of
possible board games. This gives a tree-like or hierarchical
structure of reality. Reality is experienced as we look down the
tree while `awareness' or enlightenment is achieved as we look up
the tree. When we are born we take on millions and millions of
assumptions or rules through communication, which creates our day
to day perception of reality, we then spend large parts of our
lives questioning and attempting to unconstruct these assumptions
as we seek understanding of the world.

Ten years ago I would have had to apologise to the reader for
presenting such a philosophy or `metamodel' of physics but, as
mentioned in the Introduction, now that theories of everything are
beginning to be bandied about I do believe it is time to give
deeper thought to these issues. As a matter of fact the paper
\cite{Ma:pri} on which most of Section~2 is based was submitted in
1987 to the Canadian {\em Philosophy of Science Journal} where a
very enthusiastic referee conditionally accepted the paper but
insisted that the arguments were basically Kantian and that I had
to read Kant.\footnote{I duly spent the entire summer of 1989
reading up Kant and revising the paper; after which the referee
rejected the paper with the immortal words `now that the basic
structure of the author's case is more exposed I do not find it
clarified'!} Kant basically said that reality was a product of
human thought. From this perspective the fact that life appears
somewhere near the middle of Figure~3, apart from the obvious
explanation that phenomena become simpler as we approach the
boundaries hence most complex in the middle so this is
statistically where life would develop, has a different
explanation: we created our picture of physical reality around
ourselves and so not surprisingly we are near the middle.

\section{Fourier theory}

It is now high time to turn from philosophy to more mathematical
considerations. We give more details about Fourier duality and in
particular how it leads to quantum groups as a concrete `toy
model' setting to explore the above ideas. At the same time it
should be clear from the general nature of the discussion above
that quantum groups and even noncommutative geometry itself are
only relatively simple manifestations of even more general ideas
that might be approached along broadly similar lines.

First of all, usual Fourier theory on $\R$ is a pairing of two groups,
position $x$ and momentum $p$. The momentum here labels the
characters on $\R$, i.e the elements of the dual group $\hat \R$.
The corresponding character is the plane wave
\[ \chi_p(x)=e^{\imath xp}\]
The group $\hat \R$ has its group structure given by pointwise multiplication
\[ \chi_p\chi_{p'}(x)=\chi_p(x)\chi_{p'}(x)=\chi_{p+p'}(x)\]
which is therefore isomorphic to $\R$ as the addition of momentum. Moreover,
the situation is symmetrical i.e. one could regard the same plane waves
as characters $\chi_x(p)$ on momentum space. The Fourier transform is a
map from functions on $\R$ to functions on $\hat \R$,
\[ \CF(f)(p)=\int dx f(x)\chi_p(x)\]

\subsection{Loop variables and Fourier duality}

It is well-known that these ideas work for any locally compact
Abelian group. The local-compactness is needed for the existence
of a translation-invariant measure. As physicists we can also
apply these ideas formally for other groups pretending that there
is such a measure. For example in \cite{Ma:aim}\cite{Ma:pho}\cite{Ma:fou}\cite{Ma:csta}
we proposed a Fourier theory approach to the quantisation of photons as follows.
The elements $\kappa$ of the group are disjoint unions of oriented
knots (i.e. links) with a product law that consists of erasing any
overlapping segments of opposite orientation. The dual group is
$\CA/\CG$ of  $U(1)$ bundles and (distributional)
connections $A$ on them. Thus given any bundle and connection, the
character is the holonomy
\[ \chi_A(\kappa)=e^{\imath\int_\kappa A}.\]

We considered this set-up in and the inverse Fourier transform of
some well-known functions on $\CA/\CG$ as functions on the group of knots.
For example\cite{Ma:fou},
\eqn{foucs}{ \CF^{-1}({\rm CS})(\kappa)=\int \extd A \ {\rm CS(A)}e^{-
\imath\int_\kappa A}
=e^{\frac{\imath}{2\alpha}{\rm link}(\kappa,\kappa)}}
\eqn{foumax}{ \CF^{-1}({\rm Max})(\kappa)=\int \extd A \ {\rm Max(A)}e^{-\imath\int_\kappa A}
=e^{\frac{\imath}{2\beta}{\rm ind}(\kappa,\kappa)}}
where
\[ {\rm CS}(A)=e^{\frac{\alpha\imath}{2}\int A\wedge \extd A},\quad {\rm Max}(A)
=e^{\frac{\beta\imath}{2}\int {}^*\extd A\wedge\extd A}\]
are the Chern-Simmons and Maxwell actions, $\rm link$ denotes linking
number, $\rm ind$ denotes mutual inductance.

The diagonal $\rm ind(\kappa,\kappa)$ is the {\em mutual self-inductance} i.e. you can literally cut the knot, put
a capacitor and measure the resonant frequency to measure it. By the way,
to make sense of this one has to use a wire of a finite thickness --
the self-inductance has a log divergence. This is also the log-divergence
of Maxwell theory when one tries to make sense of the functional integral,
i.e. renormalisation has a clear physical meaning in this context\cite{Ma:fou}.

Meanwhile, $\rm link(\kappa,\kappa)$ is the self-linking number\cite{Ma:aim}\cite{Ma:pho}\cite{Ma:fou}
 of a knot
with itself, defined as follows. First of all, between
two disjoint knots $\link(\kappa,\kappa')$ is the linking number as
usual. We then introduce the following
{\em regularised linking number}
\[ {\rm link}_\eps(\kappa,\kappa')=\int_{||\vec\eps||<\eps}\extd^3\vec\eps\ \link
(\kappa,\kappa'_{\vec \eps})\]
where $\kappa'_{\vec \eps}$ is the knot displaced by the vector $\vec\eps$. The
integrand is defined almost everywhere and hence integrable. Finally, we define
the linking number as the limit of this as $\eps\to 0$, which is now defined
even when knots touch or even on the same knot. At the time of
\cite{Ma:aim}\cite{Ma:pho}\cite{Ma:fou}, actually back in 1986, I made the following conjecture which is still open.

\begin{conjecture}
Intersections that are worse and worse (i.e. so that higher and higher
derivatives coincide at the point of intersection) contribute fractions with
greater and greater denominators to the regularised linking number, but the
linking number remains in $\Q$. In the extreme
limit of total overlap the self-linking number is a generic element of $\R$.
\end{conjecture}

As evidence, if the knots intersect transversally then it is easy to see that one
obtains for the regularised linking an integer $\pm \frac{1}{2}$. This is just
because half the displacements will move one knot in to link more with the other, and
the other half to unlink. \footnote{This result for transverse intersections, the
regularised linking itself from the conjecture for higher intersections were shown to Abbay
Ashtekar (and Lou Kauffman) during the ICAMP meeting in Swansea 1988 in advance of
the eventual publication in \cite{Ma:fou}.} Although the conjecture remains open, it
does appear that it could be interesting for
loop variable quantum gravity where it would imply certain rationality
properties. By the way, one might need to average over infinitesimal
rotations as well as displacements to prove it.

Note also that our point of view in \cite{Ma:aim}\cite{Ma:pho}\cite{Ma:fou} was {\em distributional} because
as well as considering honest smooth connections we considered `connections'
defined entirely by their holonomy. In particular, given a knot $\kappa$
we defined the distribution $A_\kappa$ by its character as
\[ e^{\imath\int_{\kappa'}A_\kappa}=e^{\imath{\rm link}(\kappa,\kappa')}.\]
Such distributions are quite interesting. For example\cite{Ma:pho}\cite{Ma:csta}
if one formally evaluates the Maxwell action in these one has\cite{Ma:aim}
\eqn{string}{ {\rm Max}(A_\kappa)=e^{\frac{\imath}{4\beta}\delta^2(0)\int_\kappa \extd t \dot\kappa\cdot\dot\kappa},}
the Polyakov string action. In other words, string theory can be embedded into
Maxwell theory by constraining the functional integral to such `vortex'
configurations. An additional Chern-Simons term becomes similarly a `topological mass term'
$\link(\kappa,\kappa)$ that we proposed to be added to the Polyakov action.

Finally, these ideas also have analogues in the Hamiltonian formulation.
Thus the CCR's for the gauge field can be equivalently formulated
as
\[ [\int_\kappa A,\int_{\Sigma}E]=4\pi\imath\alpha{\rm link}(\kappa,\del\Sigma)\]
which is a signed sum of the points of intersection of the loop
with the surface. This is the point of view by which loop
variables were introduced in physics in the 1970's (as an approach
to QCD on lattices) by Mandelstam and others. We have observed in \cite{Ma:aim} that
this has an interpretation as noncommutative geometry, generalising
the noncommutative torus $v^nu^m=e^{\imath\alpha mn}u^mv^n$ to
\eqn{nonpho}{
v_\kappa u_{\kappa'}=e^{4\pi\imath\alpha{\rm
link}(\kappa,\kappa')}u_{\kappa'}v_\kappa}
 where integers are
replaced by knots or links. Here the physical picture is
\eqn{uvAB}{ u_\kappa=e^{\imath\int_\kappa A},\quad
v_\kappa=e^{\imath\int_\kappa \tilde A}} where $\tilde A$ is a dual connection
such that $E=\extd \tilde A$. So constructing the $u,v$ is equivalent to
constructing some distributional operators $A,E$ with the usual CCR's. This point of
view from \cite{Ma:aim}\cite{Ma:pho} was eventually published in
\cite{Ma:csta} as a noncommutative-geometric approach to the quantisation
of photons.

It is also an interesting question how all of these ideas
generalise from $U(1)$ to nonAbelian groups. Thus, in place of the
Abelian group of knots one can first of all consider some kind of
nonAbelian group of parameterized loops in the manifold, i.e. maps
rather than the images of these maps. (The inequivalent classes of
elements in this are the fundamental group $\pi_1$ of the
manifold.) This should be paired via the Wilson loop or holonomy
with nonAbelian bundles and connections. The precise groups and
their duality here is a little hazy but one should think of this
roughly speaking as what goes on in the construction of knot
invariants from the WZW model (or from quantum group). Thus one
could argue\cite{Ma:pho}\cite{Ma:fou} that the relationship
between the Jones polynomial $J$ and $SU_2$-Chern-Simons theory
should be viewed as {\em some kind of nonAbelian Fourier
transform}
\eqn{knotjones}{ {\CF}^{-1}({\rm CS}_{SU_2})(\kappa)\sim e^{ J(\kappa)}}
with the Jones polynomial in the role of self-linking number\footnote
{This conjecture dates from 1986 at the time of \cite{Ma:aim} but was not
published until \cite{Ma:fou}, following Witten's discovery of the relation
between the WZW model and the Jones polynomial at the ICAMP in Swansea in
1988}. We will discuss Fourier transform on nonAbelian groups in the next section
using quantum group methods, though I should say that it still remains to make
(\ref{knotjones}) precise
along such lines. The reformulation of quantum group invariants as Vassiliev
invariants and the Kontsevich integrals (which generalise the linking number) could be viewed, however, as a perturbative step in this direction.

It does seems that many of these ideas have emerged in modern
times in the loop variable approach to quantum
gravity\cite{Ash:for}\cite{RovSmo:kno}, with the nonAbelian group $SU_2$ (or
another group) in place of $U(1)$. However, I want to close this
section with some ideas in this area that I still did not see
emerge. Indeed, what the loop variable approach tells us is that
the gravitational field when recast as a spin connection is in
some sense the conjugate variable to something of manifest
topological and diffeomorphism-invariant meaning -- knots and
links in the manifold. In the same spirit it is obvious that
scaler fields correspond to points in the
manifold\cite{Ma:pho}\cite{Ma:csta}. What about in the other
direction? I would conjecture that there is another field or force
in nature (possibly as yet undiscovered) corresponding to surfaces
rather than loops (and so on). Then just as gauge fields tend to
detect $\pi_1$, the new field would for example detect $\pi_2$.
Note that in the $U(1)$ case the pairing of surfaces is of course
with 2-forms (and the 2nd cohomology is the Abelianisation of
$\pi_2$) -- we would need a nonAbelian version of that.

Actually this conjecture was one of my main motivations back in
1986 in the slightly different context of a search for such
Fourier transform or `surface transport' methods for QCD. First of
all, one can ask: if the Fourier transform of the nonAbelian
Chern-Simons theory gives the quantum group link invariants as in
(\ref{knotjones}), {\em what is the Fourier transform of the
Yang-Mills action?} According to (\ref{foumax}) it should be some
kind of
 some kind of `nonAbelian self-inductance'. The extra ingredient in QCD is
of course
 confinement. Related to this is the need for some kind of `nonAbelian Stokes theorem'.
While no continuum version of the latter exists, let us suppose that is has somehow been
defined, i.e. the Lie group $G$-valued `parallel transport' of a nonAbelian Lie-algebra valued 2-form $F$
over a surface such that if $F$ is the curvature of a gauge field then
\eqn{nonstokes}{  e^{\imath\int_\Sigma F}=e^{\imath\int_{\del \Sigma} A}.}
While this is not really possible (except rather artificially on a lattice by specifying
paths parallel transporting back to a fixed based point) we suppose something {\em like}
this.

\begin{conjecture}
With such a nonAbelian surface transport, the  QCD  vacuum expectation value of the
flux of the quantized curvature $F$ through a closed surface is an invariant of the surface.
\end{conjecture}

The point is that one usually considers only planar spans of loops in QCD
and  Wilson's criterion for confinement
says that these are area law. On the other hand if one considered a small planar loop
spanned by a large surface `ballooning out' from the loop
one would still expect some finite result (since a large area), but on the other
hand the boundary curve itself could be shrunk to zero  so that
its planar spanning surface also shrinks to zero and Wilson's criterion would give 1. The conjecture
 is that these two effects cancel out and one has in fact something
that depends only on the topological class of the surface. This does require, however,
making sense of (\ref{nonstokes}) which might require some accompanying
new fields. On the other hand, at least one standard objection to the above
ideas {\em was} solved, namely we do not need to take traces of the holonomies etc., which means that
we are considering the expectations of gauge-non-invariant
operators. It was argued in \cite{Ma:gau} that one could do this in the
context of a version of the background field method. This is important because
one can then analyse and prove confinement locally as the statement that
the expectation $<F>$ is a (nonAbelian) curvature + a non-curvature part
(the latter was shown in \cite{Ma:gau} to be the skew-symmetrized gluon
two-point function). The first part is `perimeter law' and the second is `area law'
and corresponds to confinement infinitesimally.  The conjecture would
extend these ideas globally. At the end of the day, however, the strong force
itself might emerge as related to surfaces in much the same way as gravity is to loops
via the loop gravity and spin connection formalisms.

\subsection{NonAbelian Fourier Transform}

To generalise Fourier theory beyond Abelian groups we really have
to pass to the next more general self-dual category, which is that
of Hopf algebras or quantum groups. A Hopf algebra is

\begin{itemize}

\item A unital algebra $H,1$ over the field $\C$ (say)

\item A coproduct $\Delta:H\to H\tens H$ and counit $\eps:H\to \C$
forming a {\em coalgebra}, with $\Delta,\eps$ algebra
homomorphisms.

\item An antipode $S:H\to H$ such that
$\cdot(S\tens\id)\Delta=1\eps=\cdot(\id\tens S)\Delta$.
\end{itemize}

Here a coalgebra is just like an algebra but with the axioms
written as maps and arrows on the maps reversed. Thus
coassociativity means
\eqn{coassoc}{ (\Delta\tens\id)\Delta=(\id\tens\Delta)\Delta}
etc. The axioms mean that the adjoint maps $\Delta^*:H^*\tens
H^*\to H^*$ and $\eps^*:\C\to H^*$ make $H^*$ into an algebra.
Here $\eps^*$ is simply $\eps$ regarded as an element of $H^*$.
The meaning of the antipode $S$ is harder to explain but it
generalises the notion of inverse. It is a kind of `linearised
inversion'.

For a Hopf algebra, at least in the finite-dimensional case (i.e.
with a suitable definition of dual space in general) the axioms are such that
$H^*$ is again a Hopf algebra. Its coproduct is the adjoint of the
product of $H$ and its counit is the unit of $H$ regarded as a map
on $H^*$. This is why the category of Hopf algebras is a self-dual
one. For more details we refer to \cite{Ma:book}.

We will give examples in a moment, but basically these axioms are
set up to define Fourier theory. Thinking of $H$ as like
`functions on a group', the coproduct corresponds to the group
product law by dualisation. Hence a translation-invariant integral
means in general a map  $\int:H\to \C$ such that
\eqn{int}{ (\int\tens\id)\Delta=1\int }
Meanwhile, the notion of plane wave or exponential should be
replaced by the canonical element
\eqn{exp}{ \exp=\sum_a e_a\tens f^a\in H\tens H^*}
where $\{e_a\}$ is a basis and $\{f^a\}$ is a dual basis. We can
then define Fourier transform as
\eqn{fou}{ \CF:H\to H^*,\quad \CF(h)=\int (\exp) h=(\int\sum_a e_a h)f^a.}
There is a similar formula for the inverse $H^*\to H$.

The best way to justify all this is to see how it works on our
basic example for Fourier theory. Thus, we take $H=\C[x]$ the
algebra of polynomials
in one variable, as the coordinate algebra of $\R$. It forms a
Hopf algebra with
\eqn{Rcoprod}{\Delta x=x\tens 1+1\tens x, \quad \eps x=0\,\quad Sx=-x}
as an expression of the additive group structure on $\R$.
Similarly we take $\C[p]$
for the coordinate algebra of another copy of $\R$ with generator $p$
dual to $x$ (the additive group $\R$ is self-dual).

\begin{example} The Hopf algebras $H=\C[x]$ and $H^*=\C[p]$ are dual to
each other with $\<x^n,p^m\>=(-\imath)^n\delta_{n,m}n!$ (under which the coproduct of
one is dual to the product of the other). The exp element and Fourier
transform is therefore
\[ \exp=\sum\imath^n\frac{x^n\tens p^n}{n!}=e^{\imath x\tens p},\quad \CF(f)(p)=\int_{-\infty}^\infty \extd x
f(x)e^{\imath x\tens p}.\]
\end{example}

Apart from an implicit $\tens$ symbol which one does not usually
write, we recover usual Fourier theory. Both the notion of duality
and the exponential series are being treated a bit formally but can
be made precise.

Let is now apply this formalism to Fourier theory on classical but nonAbelian
groups. We use Hopf algebra methods because Hopf algebras include both
groups and group duals even in the nonAbelian case, as we have promised
in Section~2. Thus, if $\cg$ is a Lie algebra with associated Lie group $G$,
we have two Hopf algebras, dual to each other. One is $U(\cg)$ the enveloping
algebra with
\[ \Delta\xi=\xi\tens 1+1\tens\xi,\quad \xi\in \cg\]
and the other is the algebra of coordinate functions $\C(G)$. If $G$ is a matrix
group the functions $t_{ij}$ which assign to a group element its $ij$ matrix
entry generate the coordinate algebra. Of course, they commute i.e.
$\C(G)$ is the commutative polynomials in the $t^i{}_j$ modulo some other
relations that characterise the group. Their coproduct
is
\[ \Delta t^i{}_j=t^i{}_k\tens t^k{}_j\]
corresponding to the matrix multiplication or group law. The pairing is
\[ \<t^i{}_j,\xi\>=\rho(\xi)^i{}_j\]
where $\rho$ is the corresponding matrix representation of the Lie
algebra. The canonical element or $\exp$ is given by choosing a
basis for $U(\cg)$ and finding its dual basis.

\begin{example} $H=\C(SU_2)=\C[a,b,c,d]$ modulo the relation $ad-bc=1$
(and unitarity properties). It has coproduct
\[ \Delta a=a\tens a+b\tens c,\quad {\rm etc.},\quad \Delta
\left(\begin{matrix}a & b\\ c &d\end{matrix}\right)
= \left(\begin{matrix}a & b\\ c &d\end{matrix}\right)\tens
\left(\begin{matrix}a & b\\ c &d\end{matrix}\right).\]
It is dually paired with $H^*=U(su_2)$ in its antihermitian usual generators $\{e_i\}$ with
pairing
\[ \<\left(\begin{matrix}a & b\\ c &d\end{matrix}\right),e_i\>
=\frac{\imath}{2}\sigma_i,\]
defined by the Pauli matrices. Let
$\{e_1^ae_2^be_3^c\}$ be a basis
of $U(su_2)$ and $\{f^{a,b,c}\}$ the dual basis. Then
\[ \exp=\sum_{a,b,c}f^{a,b,c}\tens e_1^ae_2^be_3^c\in
\C(SU_2)\bar{\tens} U(su_2)\]
\[ \CF(f)=\int_{SU_2}\extd u f(u)f^{a,b,c}(u)\tens e_1^a e_2^b e_3^c.\]
\end{example}

Here $\extd u$ denotes the right-invariant Haar measure on $SU_2$.
For a geometric picture one should think of $e_i$ as noncommuting
coordinates i.e. regard $U(su_2)$ as a `noncommutative space' as
in (\ref{su2x}). An even simpler example is the Lie algebra
$\cb_+$ with generators $x,t$ and relations $[x,t]=\imath\lambda
x$. Its enveloping algebra could be viewed as a noncommutative
analogue of 1+1 dimensional space-time.

\begin{example}c.f. \cite{MaOec:twi} The group $B_+$ of matrices of the form
\[\left(\begin{matrix}e^{\lambda\omega} & k\\ 0 &1\end{matrix}\right)\]
has coordinate algebra $\C(B_+)=\C[k,\omega]$ with coproduct
\[ \Delta e^{\lambda\omega}=e^{\lambda\omega}\tens e^{\lambda\omega},
\quad\Delta k=k\tens 1+e^{\lambda\omega}\tens k \]
Its duality pairing with $U(\cb_+)$ is generated by $\<x,k\>=-\imath, \<t,\omega\>=-\imath$
and the resulting exp and Fourier transform
are
\[ \exp=e^{\imath k\omega}e^{\imath\omega t},\quad \CF(:f(x,t):)
=\int_{-\infty}^\infty\int_{-\infty}^\infty \extd x
\extd t\ e^{\imath k x}e^{\imath\omega t}f(e^{\lambda\omega}x,t)\]
where $:f(x,t):\in U(\cb_+)$ by normal ordering $x$ to the left of $t$.
\end{example}

Similarly (putting a vector $\vec x$ in place of $x$) the algebra
$[\vec x,t]=\imath\lambda \vec x$ is merely the enveloping algebra
of the Lie algebra of the group $\R\rcross\R^n$ introduced (for
$n=2$) in \cite{Ma:mat} and could be viewed as some kind of
noncommutative space-time in $1+n$ dimensions. This was justified
in 1+3 dimensions in \cite{MaRue:bic}, where it was shown to be
the correct `kappa-deformed' Minkowski space covariant under a
`kappa-deformed' Poincar\'e quantum group which had been proposed
earlier\cite{LNRT:def}. We see that Fourier transform then
connects it to the classical coordinate algebra
$\C(\R\rcross\R^n)$ of the nonAbelian group $\R\rcross\R^n$, this
time with commuting coordinates $(\vec k,\omega)$. This
demonstrates in detail what we promised that noncommutavity of
spacetime is related under Fourier transform to nonAbelianness
(which typically means curvature) of the momentum group. Under
Fourier theory it means that {\em all} noncommutative geometrical
constructions and problems on this spacetime can be mapped over
and solved as classical geometrical constructions on the
nonAbelian momentum space.

This Fourier transform approach was demonstrated recently in
\cite{AmeMa:wav}, where we analyse the gamma-ray burst experiments
mentioned in Giovanni Amelino-Camelia's lectures at the
conference, from this  point of view. In contrast to
previous suggestions\cite{ALN:def} (based on the deformed Poincar\'e
algebra) we are able to justify the dispersion relation
\eqn{disp}{ \lambda^{-2}\left(e^{\lambda\omega}+e^{-\lambda\omega}-2\right)
-\vec k^2 e^{-\lambda\omega}=m^2}
as a well-defined mass-shell in the classical momentum group $\R\rcross\R^3$
and give some arguments that the plane waves being of the form
$e^{\imath \vec k\cdot\vec x}e^{\imath \omega t}$ above would
have wave velocities given by $v_i=\frac{\del \omega}{\del k_i}$ (no
meaningful justification for this of any kind had been given before).
In particular, one has a variation in arrival time for a gamma ray emitted 
a distance $L$
away
\eqn{deltaT}{ \delta T=\lambda L\delta k=\lambda\frac{L}{c}\delta
\omega} as one varies the momentum by $\delta k$ or the energy
by $\delta\omega$. Apparently such
theoretical predictions can actually be measured for gamma ray
bursts that travel cosmological distances. Of course, one needs to
know the distance $L$ and use the predicted $
L$-dependence to filter out other effects and also to filter out
our lack of knowledge of the initial spectrum of the bursts. It is
also conjectured in \cite{AmeMa:wav} that the nonAbelianness of
the momentum group shows up as CPT violation and might be detected
by ongoing neutral-kaon system experiments. Of course, there is
nothing stopping one doing field theory in the form of Feynman
rules on our classical momentum group either, except that one has
to make sense of the meaning of nonAbelianess in the addition of
momentum. As explained in Section~2 one can use similar techniques
to those for working on curved position space, but now in momentum
space, i.e. I would personally call such effects, if detected,
`cogravity'. The idea is that quantum gravity should lead to both
gravitational and these more novel cogravitational effects at the
macroscopic level.

Let us note finally that these nonAbelian Fourier transform ideas also
work fine for finite groups and could be useful for crystallography.

\section{Bicrossproduct model of Planck-scale physics}

So far we have only really considered groups or their duals,
albeit nonAbelian ones. The whole point of Hopf algebras, however,
is that there exist examples going truly beyond these but with
many of the same features, i.e. with properties of groups and
group duals unified. It is high time to give some examples of Hopf
algebras going beyond groups and group duals i.e. neither
commutative like $\C(G)$ not the dual concept (cocommutative) like
$U(\cg)$, i.e. {\em genuine} quantum groups.

We recall from Section~2 that the unification of groups and group
duals is a kind of microcosm or `toy model' of the problem of
unifying quantum theory and gravity. So our first class of quantum
groups (the other to be described in a later section) come from
precisely this point of view.

\subsection{The Planck-scale quantum group}

By `toy model' we mean of course some kind of effective theory
with stripped-down degrees of freedom but incorporating the idea
that Planck scale effects would show up when we try to unify
quantum mechanics and geometry through noncommutative geometry.
But actually our approach can make a much stronger statement than
this: we envisage that the model appears as some effective limit
of an unknown theory of quantum-gravity which to lowest order
would appear as spacetime and conventional mechanics on it
-- but even if the theory is unknown we can use the intrinsic
structure of noncommutative algebras to classify {\em a priori}
different possibilities. This is much as a phenomenologist might
use knowledge of topology or cohomology to classify different {\em
a priori} possible effective Lagrangians.

Specifically, if $H_1$, $H_2$ are two quantum groups there is a
theory of the space ${\rm Ext}_0(H_1,H_2)$ of possible extensions
\[ 0\to H_1\to E\to H_2\to 0\]
by some Hopf algebra $E$ obeying certain conditions. We do not
need to go into the mathematical details here but in general one
can show that $E=H_1\bicross H_2$ a `bicrossproduct' Hopf algebra.
Suffice it to say that the conditions are `self-dual' i.e. the
dual of the above extension gives
\[ 0\to H_2^*\to E^*\to H_1^*\to 0\]
as another extension dual to the first, in keeping with our
philosophy of self-duality of the category in which we work. We
also note that by $\Ext_0$ we mean quite strong extensions. There
is also a weaker notion that admits the possibilities of cocycles
as well, which we are excluding, i.e. this is only the
topologically trivial sector in a certain nonAbelian cohomology.

\begin{theorem}
\cite{Ma:pla}\cite{Ma:phy} ${\rm Ext}_0(\C[x],\C[p])=\R\hbar\oplus
\R\grav$, i.e. the different extensions
\[ 0\to \C[x]\to ?\to\C[p]\to 0\]
of position $\C[x]$ by momentum $\C[p]$ forming a Hopf algebra are
classified by two parameters which we denote $\hbar,\grav$ and
take the form
\[ ?\isom \C[x]\bicross_{\hbar,\grav}\C[p].\]
Explicitly this 2-parameter Hopf algebra is generated by $x,p$
with the relations and coproduct
\[ [x,p]=\imath\hbar(1-e^{-\frac{x}{\grav}}),\quad \Delta x=x\tens
1+1\tens x,\quad \Delta p=p\tens e^{-\frac{x}{\grav}}+1\tens p.\]
\end{theorem}

This is called the {\em Planck scale quantum group}. It is a bit
more than just some randomly chosen deformation of the coordinate
algebra of the usual group $\R^2$ of phase space of a particle in
one dimension: in physical terms what we are saying is that if we
are given $\C[x]$ the position coordinate algebra and $\C[p]$
defined {\em a priori} as the natural momentum coordinate algebra
then {\em all possible} quantum phases spaces built from $x,p$ in
a controlled way that preserves duality ideas (Born reciprocity)
and retains the group structure of classical phase space as a
quantum group are of this form labeled by two parameters
$\hbar,\grav$. We have not put these parameters in by hand -- they
are simply the mathematical possibilities being thrown at us. {\em
In effect we are showing how one is forced to discover both
quantum and gravitational effects from certain structural
self-duality considerations.}

The only physical input here is to chose suggestive names for the two
parameters by looking at limiting cases. We also should say what
we mean by `natural momentum coordinate'. What we mean is that the
interpretation of $p$ should be fixed before hand, e.g. we
stipulate before hand that the Hamiltonian is $h=p^2/2m$ for a
particle on our quantum phase space. Then the different
commutation relations thrown up by the mathematical structure
imply different dynamics. If one wants to be more
conventional then one can define $\tilde
p=p(1-e^{-\frac{x}{\grav}})^{-1}$ with canonical commutation
relations but some nonstandard Hamiltonian,
\[ [x,\tilde p]=\imath\hbar,\quad h=\frac{\tilde
p^2}{2m}(1-e^{-\frac{x}{\grav}})^2.\] Thus our approach is
slightly unconventional but is motivated rather by the strong
principle of equivalence that from some point of view the particle
should be free. We specify $x,p$ before-hand to be in that frame
of reference and then explore their possible commutation
relations. Of course the theorem can be applied in other contexts
too whenever the meaning of $x,p$ is fixed before hand, perhaps by
other criteria.

\subsubsection{The quantum flat space $\grav\to 0$ limit} Clearly in the
domain where $x$ can be treated
as having values $>0$, i.e. for a certain class of quantum states
where the particle is confined to this region, we clearly have
flat space quantum mechanics $[x,p]=\imath\hbar$ in the limit
$\grav\to 0$.

\subsubsection{The classical $\hbar\to 0$ limit} On the other hand, as $\hbar\to 0$
we just have the commutative polynomial algebra $\C[x,p]$ with the
coalgebra shown. This is the coordinate algebra of the group $B_-$
of matrices of the form
\[ \left(\begin{matrix}e^{-\frac{x}{\grav}} & 0\\ p &1\end{matrix}\right)\]
which is therefore the classical phase space for general $\grav$
of the system.

\subsubsection{The dynamics} The meaning of the parameter $\grav$ can
be identified, at least roughly, as follows. In fact the meaning
of $p$ mathematically in the construction is that it acts on the
position $\R$ inducing a flow. For such dynamical systems the
Hamiltonian is indeed naturally $h=p^2/m$ and implies that
\[ \dot p=0,\quad \dot
x=\frac{p}{m}\left(1-e^{-\frac{x}{\grav}}\right)+O(\hbar)
=v_\infty\left(1-\frac{1}{1+\frac{x}{\grav}+\cdots}\right)+O(\hbar)\]
where we identify $p/m$ to $O(\hbar)$ as the velocity $v_\infty<0$
at $x=\infty$. We see that as the particle approaches the origin
it goes more and more slowly and in fact takes an infinite amount
of time to reach the origin. Compare with the formula in standard
radial infalling coordinates
\[ \dot x=v_\infty\left(1-\frac{1}{1+\frac{1}{2}\frac{x}{\grav}}\right)\]
for the distance from the event horizon of a Schwarzschild black
hole with
\[ \grav= \frac{G_{\rm Newton}M}{c^2},\]
where $M$ is the background gravitational mass and $c$ is the
speed of light. Thus the heuristic meaning of $\grav$ in our model
is that it measures the background mass or
radius of curvature of the classical geometry of which our Planck
scale Hopf algebra is a quantisation.

These arguments are from \cite{Ma:pla}. Working a little harder,
one finds that the quantum mechanical limit is valid (the effects
of $\grav$ do not show up within one Compton wavelength) if
\[ mM<<m^2_{\rm Planck},\]
while the curved classical limit is valid
if
\[ mM>>m^2_{\rm Planck}.\]
See also \cite{Ma:book}. {\em The Planck-scale quantum group therefore
truly unifies quantum effects and `gravitational' effects
in the context of Figure~3.}

Of course our model is only a toy model and one cannot draw too
many conclusions given that our treatment is not even
relativistic. The similarity to the Schwarzschild black-hole is, however, quite
striking and one could envisage more complex examples which hit
that exactly on the nose. The best we can say at the moment is
that the search to unify quantum theory and gravity using such
methods leads to tight constraints and features such as
event-horizon-like coordinate singularities. Theorem~1 says that
it is not possible to make a Hopf algebra for $x,p$ with the correct
classical limit in this context without such a coordinate
singularity.

\subsubsection{The quantum-gravity $\hbar,\grav\to \infty, \frac{\grav}{\hbar}=\lambda$ limit}
Having analysed the two familiar limits we can consider other
`deep quantum-gravity' limits. For example sending both our
constants to $\infty$ but preserving their ratio we have
\[ [x,p]=\imath\lambda x,\quad \Delta x=x\tens 1+1\tens x,\quad
\Delta p\tens 1+1\tens p\]
which is once again $U(\cb_+)$ regarded as in Example~3 in Section~3 `up side
down' as a quantum space. The higher-dimensional analogues are
`$\kappa$-deformed' Minkowski space\cite{MaRue:bic} as explained in Section~3,
i.e. the Planck-scale
quantum group puts some flesh on the idea that this might indeed come out
of quantum gravity as some kind of effective limit\cite{MaOec:twi}. Time itself
would have to appear as $t=p$, (or $t=\sum_i p_i$
for the higher dimensional analogues) in this limit from the
 momenta conjugate in the effective quantum gravity theory to the position
coordinates. This speculative possibility is discussed further in
\cite{AmeMa:wav}. At any rate this deformed Minkowski space is at
least mathematically nothing but a special limit of the
Planck-scale quantum group from \cite{Ma:pla}. It gives some idea
how the self-duality ideas of Section~2 might ultimately connect
to testable predictions for Planck scale physics e.g. testable by
gamma-ray bursts of cosmological origin.

\subsubsection{The algebraic structure and Mach's Principle} The
notation $\C[x]\bicross \C[p]$ for the Planck-scale quantum group
reflects its algebraic structure. As an algebra it is a cross
product $\C[x]\lcross\C[p]$ by the action $\la$ of $\C[p]$ on $\C[x]$ by
\eqn{pact}{ p\la f(x)=-\imath\hbar(1-e^{-\frac{x}{\grav}})\frac{\del}{\del
x} f}
which means that it is a more or less standard `Mackey
quantisation' as a dynamical system. It can also be viewed as the
deformation-quantization of a certain Poisson bracket structure on
$\C(B_-)$ if one prefers that point of view. On the other hand its
coproduct is obtained in a similar but dual way as a semidirect
coproduct $\C[x]\rcocross\C[p]$ by a coaction of $\C[x]$ on
$\C[p]$. This coaction is induced by an action of $x$ on functions $f(p)$
of similar form to the above but with the roles reversed.
In other words, {\em matching the action of momentum on
position is an `equal and opposite' coaction of position back on
momentum}. This is indeed inspired by the ideas of Mach\cite{Ma:phy} as
was promised in Section~2.

\subsubsection{Observable-state duality and T-duality}
The phrase `equal and opposite' has a precise consequence here. Namely
the algebra corresponding to the coalgebra by dualisation has a
similar cross product form by an analogous action of $x$ on $p$. More
precisely, one can show that
\eqn{planckdual}{ (\C[x]\bicross_{\hbar,\grav}\C[p])^*\isom \C[\bar
p]\cobicross_{\frac{1}{\hbar},\frac{\grav}{\hbar}}\C[\bar x],}
where $\C[p]^*=\C[\bar x]$ and $\C[x]^*=\C[\bar p]$ in the sense
of an algebraic pairing as in Example~1 in Section~3. Here $\<p,\bar x\>
=\imath$ etc., which then requires a change of the parameters as shown to
make the identification precise. So the Planck-scale quantum group
is self-dual up to change of parameters.

This means that whereas we would look for observables $a\in
\C[x]\bicross\C[p]$ as the algebra of observables and states $\phi\in
\C[\bar p]\cobicross\C[\bar x]$ as the dual linear space, with
$\phi(a)$ the expectation of $a$ in state $\phi$ (See
section~2.2), there is a dual interpretation whereby
\[{\rm Expectation}=\phi(a)=a(\phi)\]
for the expectation of $\phi$ in `state' $a$ with $\C[\bar
p]\cobicross\C[\bar x]$ the algebra of observables in the dual theory.
More precisely, only self-adjoint elements of the algebra are
observables and positive functionals states, an a state $\phi$
will not be exactly hermitian in the dual theory etc. But the
physical hermitian elements in the dual theory will be given by
combinations of such states, and vice versa. This is a concrete
example of observable-state duality as promised in Section~2. It
was introduced by the author in \cite{Ma:pla}.

Also conjectured at the time of \cite{Ma:pla} was that this
duality should be related to $T$-duality in string theory. As
evidence is the inversion of the constant $\hbar$. In general
terms coupling inversions are indicative of such dualities. Notice
also that Fourier transform implements this T-duality-like
transformation as
\[ \CF: \C[x]\bicross_{\hbar,\grav}\C[p]\to \C[\bar
p]\cobicross_{\frac{1}{\hbar},\frac{\grav}{\hbar}}\C[\bar x]\]
Explicitly, it comes out as\cite{MaOec:twi}
\eqn{planckfou}{ \CF(:f(x,p):)=\int_{-\infty}^\infty\int_{-\infty}^\infty \extd x\extd p
\ e^{-\imath(\bar p+\frac{\imath}{\grav})x}e^{-\imath\bar x(p+p\la)}
f(x,p),}
where $\la$ is the action (\ref{pact}) and $f(x,p)$ is a
classical function considered as defining an element of the
Planck-scale quantum group by normal ordering $x$ to the left.

The duality here is not exactly T-duality in string theory but has
some features like it. On the other hand it is done here at the
quantum level and not in terms only of Lagrangians. In this sense
the observable state duality can give an idea about what should be
`M-theory' in string theory. Thus, at the moment all that one
knows really is that the conjectured M-theory should be some form
of algebraic structure with the property that it has different
semiclassical limits with different Lagrangians related to each
other by S,T dualities (etc.) at the classical level. Our observable-state
duality ideas \cite{Ma:pla}\cite{Ma:pri}\cite{Ma:som} as well as
more recent work on T-duality suggests that:

\begin{conjecture} M-theory should be some kind
of algebraic structure possessing one or more dualities in a
representation-theoretic or observable-state sense.
\end{conjecture}

Actually there is an interesting anecdote here. I once had a chance
to explain the algebraic duality ideas of my PhD thesis to Edward
Witten at a reception in MIT in 1988 after his colloquium talk at Harvard on
the state of string theory. He asked me `is there a
Lagrangian' and when I said `No, it is all algebraic; classical
mechanics only emerges in the limits, but there are two different
limits related by duality', Witten rightly (at the time) gave me
a short lecture about the need for a Lagrangian.
9 years later I was visiting Harvard and Witten gave a similarly-titled
colloquium talk on the
state of string theory. He began by stating that there was some
algebraic structure called M-theory with Lagrangians appearing
only in different limits.

\subsubsection{The noncommutative differential geometry}
The lack of Lagrangians and other familiar structures in the full
Planck-scale theory was certainly a valid criticism back in
1988. Since then, however, noncommutative geometry has come a long
way and one is able to `follow' the geometry as we quantise the
system using these modern techniques. We do not have the space to
recall the whole framework but exterior algebras, partial
derivatives etc., make sense for quantum groups and many other
noncommutative geometries. For the Planck-scale quantum group one
has\cite{MaOec:twi},
\eqn{planckdifa}{ \del_p:f(x,p):=\frac{\grav}{\imath\hbar}:(f(x,p)-f(x,p-\imath\frac{\hbar}{\grav})):,}
\eqn{planckdifb}{ \del_x:f(x,p):=:\frac{\del}{\del x} f:-\frac{p}{\grav}\del_p:f:}
which shows the effects of $\hbar$
in modifying the geometry. Differentiation in the $p$ direction
becomes `lattice regularised' albeit a little strangely with an
imaginary displacement. In the deformed-Minkowski space setting
where $p=t$ it means that the Euclidean version of the theory
is related to the Minkowski one by a Wick-rotation is being
lattice-regularised by the effects of $\hbar$.

Also note that for fixed $\hbar$ the geometrical picture blows up
when $\grav\to 0$. I.e the usual flat space quantum mechanics CCR
algebra does not admit a deformation of conventional differential
calculus on $\R^2$ -- {\em one needs a small amount of `gravity'
to be present for a geometrical picture in the quantum theory.}
This is also evident in the exterior algebra\cite{MaOec:twi}
\[ f\extd x=(\extd x)f,\quad f\extd p=(\extd
p)f+\frac{\imath\hbar}{\grav}\extd f\] for the relations between
`functions' $f$ in the Planck-scale quantum group and
differentials. The higher exterior algebra looks more innocent
with
\eqn{planckext}{
\extd x\wedge
\extd x=0,\quad \extd x\wedge \extd p=-\extd p\wedge\extd x,\quad
\extd p\wedge \extd p=0}
Starting with the differential forms and derivatives, one can
proceed to gauge theory, Riemannian structures etc., in some
generality. One can also write down `quantum' Poisson brackets and
Hamiltonians\cite{MaOec:twi} and (in principle) Lagrangians in the
full noncommutative theory. Such tools should help to bridge the
gap between model building via classical Lagrangians, which I
personally do not think can succeed at the Planck scale, and
some of the more noncommutative-algebraic ideas in Section~2.

\subsection{Higher dimensional analogue}

The Planck-scale quantum group is but the simplest in a family of
quantum groups with similar features and parameters. We work from
now with $\grav=\hbar=1$ for simplicity but one can always put the
parameters back.

Of course one may take the $n$-fold tensor product of the
Planck-scale quantum group, i.e. generators $x_i,p_i$ and
different $i$ commuting.  However, in higher dimensions the
$\rm Ext_0$ is much bigger and I do not know of any full computation
of all the possibilities for $n>1$. More interesting perhaps
 are some genuinely different higher-dimensional examples along
similar but nonAbelian lines, one of which we describe now. The material
is covered in \cite{Ma:book}, so we will be brief.

Thus, also from 1988, there is a bicrossproduct quantum
group
\eqn{3planck}{ \C(\R\rcross\R^2)\bicross U(su_2)}
constructed in \cite{Ma:mat}\cite{Ma:hop} (actually as a Hopf-von
Neumann algebra; here we consider only the simpler algebraic
structure underlying it.)

The nonAbelian group $\R\rcross\R^2$ is the one whose enveloping algebra
we have considered in Example~3 in Section~3 as noncommutative spacetime.
Here, however,
we take it with a Euclidean signature and a different notation.
Explicitly, it consists of 3-vectors $\vec s$ with third component $s_3>-1$ and with the `curved
$\R^3$' nonAbelian group law
\[ \vec{s}\cdot\vec{t}=\vec{s}+(s_3+1)\vec{t}.\]
Its Lie algebra is spanned by $x_0,x_i$ with relations $[x_i,x_0]=x_i$
for $i=1,2$ as discussed before (this is how this algebra appeared first,
in \cite{Ma:mat}, in connection this higher-dimensional version of the
Planck-scale quantum group). Now, on the group $\R\rcross\R^2$ there is an
action of $SU_2$ by a deformed rotation. This is shown in
Figure~4. The orbits are still spheres but non-concentrically
nested and accumulating at $s_3=-1$. This is a dynamical system
and (\ref{3planck}) is its Mackey quantisation as a cross
product. We see that we have similar features as for the
Planck-scale quantum group, including some kind of coordinate
singularity as $s_3=-1$.
\begin{figure}
\[ \epsfbox{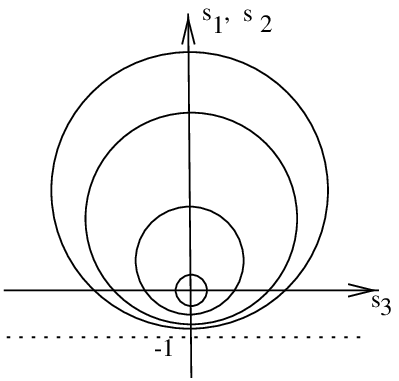}\]
\caption{Deformed action of classical $SU_2$ on $\R\rcross\R^2$}
\end{figure}

At the same time there is a `back reaction' of $\R\rcross\R^2$
back on $SU_2$, which appears as a coaction of $\C(\R\rcross\R^2)$
on $U(su_2)$ in the cross coalgebra structure of the quantum
group. Therefore the dual system, related by Fourier theory or
observable-state duality, is of the same form, namely
\eqn{dual3planck}{ U(\R\rcross\R^2)\cobicross \C(SU_2).}
It consists of a particle on $SU_2$ moving under the action of
$\R\rcross\R^2$. This is the dual system which, in the present case, looks
quite different.

Finally, the general theory of bicrossproducts allows for a
`Schroedinger representation' of (\ref{3planck}) on
$U(\R\rcross\R^2)$ and similarly of its dual on $U(su_2)$. Such a picture
means that the `wave functions' live in these enveloping algebras
viewed as noncommutative spaces. There are also more conventional
Hilbert space representations as well.

\subsection{General construction} There is a general
construction for bicrossproduct quantum groups of which the ones
discussed so far are all examples. Thus suppose that
\[ X=GM\]
is a {\em factorisation of Lie groups}. Then one can show that $G$
acts on the set of $M$ and $M$ acts back on the set of $G$ such
that $X$ is recovered as a double cross product (simultaneously by
the two acting on each other) $X\isom G\dcross M$. This turns out
to be just the data needed for the associated cross product and
 cross coproduct
\eqn{facbic}{\C(M)\bicross U(\cg)}
 to be a Hopf algebra. The roles of the two Lie groups is
 symmetric and the dual is
 \eqn{facdual}{(\C(M)\bicross U(\cg))^*=U(\cm)\cobicross \C(G)}
 which means that there are certain
 families of homogeneous spaces (the orbits of one group under
 the other) which come in pairs, with the algebra of observables
 of the quantisation of one being the algebra of expectation
 states of the quantisation of the other. This
 is the more or less purest form of the ideas of Section~2 based
 on Mach's principle\cite{Ma:phy} and duality.

On the other hand, factorisations abound in Nature. For example
every complexification of a simple Lie group factorises into its
compact real form $G$ and a certain solvable group $G^\star$, i.e.
$G_\C=GG^\star$. The notation here is of a modern approach to the Iwasawa
decomposition in \cite{Ma:mat}. For example, $SL_2(\C)=SU_2
SU_2^\star$, where $SU_2^\star=\R\rcross\R^2$, gives the
bicrossproduct quantum group (\ref{3planck}) in the preceding
section. There are similar examples
\eqn{ggstarbic}{ \C(G^\star)\bicross U(\cg)}
for all complex simple $\cg$. Also, slightly more general than the
Iwasawa decomposition but still only a very special case of a
general Lie group factorisation, let
$G$ be a Poisson-Lie group (a Lie group with a compatible
Poisson-bracket). At the infinitesimal level the Poisson bracket
defines a map $\cg\to\cg\tens \cg$ making $\cg$ into a Lie
bialgebra. This is an infinitesimal idea of a quantum group and is
such that $\cg^\star$ is also a Lie algebra. In this setting there
is a Drinfeld double Lie bialgebra $D(\cg)$ and its Lie group is
an example of a factorisation $GG^\star$.

By the way, this is
exactly the setting for nonAbelian Poisson-Lie T-duality
\cite{Kli:poi} in string
theory, for classical $\sigma$-models on $G$ and $G^\star$. The
quantum groups (\ref{ggstarbic}) and their duals are presumably
related to the quantisations of the point-particle limit of these
sigma models. If so this would truly extend T-duality to the
quantum case via the above observable-state duality ideas. While this is not
proven exactly, something {\em like} this appears to be the case.
Moreover, the bicrossproduct duality for (\ref{facbic}) is much
more general and is not limited to such Poisson-Lie structures on
$G$. The group $M$ need not be dual to $G$ in the above sense and
need not even have the same dimension. Recently it was shown that
the Poisson-Lie T-duality in a Hamiltonian (but not Lagrangian)
setting indeed generalises to a general factorisation like
this\cite{BegMa:poi}.

Finally, there is one known connection between the bicrossproduct
quantum groups and the more standard $U_q(\cg)$ which we will
consider next. Namely, Lukierski et al.\cite{LNRT:def} showed that a
certain contraction process turned $U_q(so_{3,2})$ in a certain
limit to some kind of `$\kappa$-deformed' Poincar\'e algebra as mentioned
below Example~3 in Section~3. It
turned out later\cite{MaRue:bic} that this was isomorphic to one
of the bicrossproduct Hopf algebras above,
\[ {}_\kappa{\rm Poincare}\isom \C(\R\rcross\R^3)\bicross
U(so_{3,1}).\] The isomorphism here is nontrivial (which means in
particular that $\kappa$-Poincar\'e certainly arose independently
of the early bicrossproducts such as the 3-dimensional case
(\ref{3planck})). On the other hand, the bicrossproduct version of
$\kappa$-Poincar\'e from \cite{MaRue:bic} brought many benefits.
First of all, the Lorentz sector is {\em undeformed}. Secondly,
the dual is easy to compute (being an example of the general
self-duality ideas above) and, finally, the Schroedinger
representation means that this quantum group indeed acts
covariantly on $U(\R\rcross\R^3)$, which should therefore be
viewed as the $\kappa$-Minkowski space appropriate to this
$\kappa$-Poincar\'e (prior to \cite{MaRue:bic} one had only the
noncovariant action of it on usual commutative Minkowski space,
leading to a number of inconsistencies in attempting to model
physics based on $\kappa$-Poincar\'e alone). Of course the point
of view of Poincar\'e algebra as symmetry appears at first
different from the main point of view of bicrossproducts as the
quantisations of a dynamical system. However, as in Section~2 (and
even for the classical Poincar\'e algebra) a symmetry enveloping
algebra {\em should} also appear as part of (or all of) the
quantum algebra of observables of the associated quantum theory
because it should be realised among the quantum
fields\cite{Ma:ista}.

\section{Deformed quantum enveloping algebras}

No introduction to quantum groups would be complete if we did not
also mention the much more well known deformations $U_q(\cg)$ of
complex simple $\cg$ arising from inverse scattering and the
theory of solvable lattice models\cite{Dri}\cite{Jim:dif}. These
have not, however, been very directly connected with Planck scale
physics (although there are some recent proposals for this, as we
saw in the lectures of Lee Smolin). They certainly did not arise
that way and are not the quantum algebras of observables of
physical systems. Therefore this is only going to be a lightning
introduction to this topic. For more, see
\cite{Ma:book}\cite{Ma:introp}\cite{Ma:varen}.

Rather, these quantum groups $U_q(\cg)$ arise naturally as
`generalised' symmetries of
certain spin chains and as generalised symmetries in the
Wess-Zumino-Witten model conformal field theory. Just as groups
can be found as symmetries of many different and unconnected
systems, the same is true for the quantum groups $U_q(\cg)$. They
do, however, have a perhaps richer and more complex mathematical
structure than the bicrossproducts, which is what we shall briefly
outline.

As Hopf algebras one has the same duality ideas nevertheless.
Thus, the quantum group $U_q(su_2)$ with generators $H,X_\pm$ and
relations and coproduct
\[ [H,X_\pm]=\pm X_\pm,\quad
[X_+,X_-]=\frac{q^H-q^{-H}}{q-q^{-1}}\]
\[ \Delta X_\pm=X_\pm\tens q^\frac{H}{2}+q^\frac{-H}{2}\tens
X_\pm,\quad \Delta H=H\tens 1+1\tens H\] is dual to the quantum
group $\C_q(SU_2)$ generated by a matrix of generators $a,b,c,d$.
This has six relations of $q$-commutativity
\[ ba=qab,\  ca=qac,\  bc=cb,\  dc=qcd,\
db=qbd,\  da=ad+(q-q^{-1})bc\] and a determinant relation
$ad-q^{-1}bc=1$. The pairing is the same as in Example~2 in Section~2 at
the level of generators (after a change of basis).

The main feature of these quantum groups, in contrast to the
bicrossproduct ones, is that their representations form braided
categories. Thus, if $V,W\in {\rm Rep}( U_q(\cg))$ then
$V\tens W$ is (as for any quantum group) also a representation. The action is
\eqn{hact}{ h\la(v\tens w)=(\Delta h).(v\tens w)}
for all $h\in U_q(\cg)$, where we use the coproduct (for example the linear form of the
coproduct of $H$ means that it acts additively). The special
feature of quantum groups like $U_q(\cg)$ is that there is an
element $\CR\in U_q(\cg)\bar{\tens} U_q(\cg)$ (the `universal R-matrix or
quasitriangular structure') which ensures an isomorphism of
representations by
\eqn{Psi}{ \Psi_{V,W}:V\tens W\to W\tens V,\quad \Psi_{V,W}(v\tens
w)=P\circ \CR.(v\tens w)} where $P$ is the usual permutation or
flip map. This {\em braiding} $\Psi$ behaves much like the usual
transposition or flip map for vector spaces but does not square to
one. To reflect this one writes $\Psi=\epsfbox{braid.eps}$,
$\Psi^{-1}=\epsfbox{braidinv.eps}$. It has properties consistent
with the braid relations, i.e. when two braids coincide the
compositions of $\Psi,\Psi^{-1}$ that they represent also coincide.
The fundamental braid relation of the braid group in Figure~5(a)
corresponds to the famous Yang-Baxter or braid relation for the
matrix corresponding to $\Psi$.
\begin{figure}
\[ (a)\quad\epsfbox{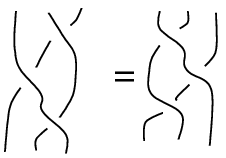}\quad (b)\epsfbox{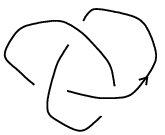}\quad(c)\ \epsfbox{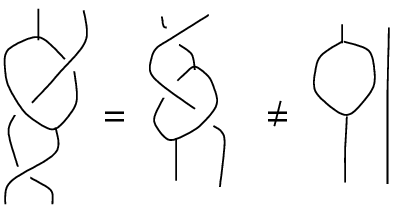}\]
\caption{(a) Braid relations (b) Trefoil knot (c) Braided algebra calculation}
\end{figure}

From this it is more or less obvious that such quantum groups lead
to knot invariants. One can scan the (oriented) knot such as in
Figure~5(b) from top to bottom. We choose a representation $V$
with dual $V^*$ and label the knot by $V$ against a downward arc
and $V^*$ against an upward arc. As we read the knot, when we
encounter an arc ${}_V\cap_{V^*}$ we let it represent the
canonical element $\sum_a e_a\tens f^a\in V\tens V^*$. When we
encounter crossings we represent them by the appropriate $\Psi$
and finally when we encounter ${}^{V^*}\cup^V$ we apply the
evaluation map. There is also a prescription for when we encounter
${}_{V^*}\cap_V$ and ${}^V\cup^{V^*}$. At the end of the day we
obtain a number depending on $q$ (which went into the braiding).
This function of $q$ is (with some fiddling that we have not
discussed) an invariant of the knot regarded as a framed knot.
This is not the place to give details of knot theory, but this is
the rough idea. In physical terms one should think of the knot as
a process in 1+1 dimensions in which a particle $V$ and
antiparticle $V^*$ is created at an arc, some kind of scattering
$\Psi$ occurs at crossings, etc.

For standard $U_q(\cg)$ the construction of representations is not
hard, all the standard ones of $\cg$ just q-deform. For example,
the spin $\h$ representation of $su_2$ deforms to a 2-dimensional
representation of $U_q(su_2)$. The associated knot invariant is
the celebrated Jones polynomial.

\subsection{Braided mathematics and braided groups}

This braiding is the key property of the quantum groups
$U_q(\cg)$ and other `quasitriangular Hopf algebras' of
similar type. It means in particular that {\em any algebra on
which the quantum group acts covariantly becomes braided}. This is
therefore indicative of a whole braided approach to noncommutative geometry
or {\em braided geometry} via algebras or `braided' spaces on which
quantum groups
$U_q(\cg)$ act as generalised symmetries. Note that we are not so
much interested in this point of view in the noncommutative
geometry of the quantum groups $U_q(\cg)$ themselves, although one
can study this as a source of mathematical examples. More physical
is the algebras in which these objects act.

In this approach the meaning of $q$ is that it enters into the
braiding, i.e. it generalises the $-1$ of supertransposition in
super-geometry. This is `orthogonal' to the usual idea of
noncommutative geometry, i.e. it is not so much a property of one
algebra but of composite systems, namely of the noncommutativity
of tensor products. The simplest new case is where the braiding is
just a factor $q$. To see how this works, consider the {\em
braided line} $B=\C[x]$. As an algebra this is just the
polynomials in one variable again.

\begin{example} Let $B=\C[x]$ be the braided line, where independent copies
$x,y$ have braid statistics $yx=qxy$ when one is transposed past
the other (c.f. a Grassmann variable but with $-1$ replaced by
$q$). Then
\[ \del_q f(y)=x^{-1}\left(f(x+y)-f(x)\right)|_{x=0}=\frac{f(y)-f(qy)}{(1-q)y}\]
\end{example}

This is easy to see on monomials, i.e. $\del_q y^n$ is the coefficient
of the $x$-linear
part in $(x+y)^n$ after we move all $x$ to the left.
In fact mathematicians have played with such a q-derivative since
1908\cite{Jac:fun} as having many cute properties. We see\cite{Ma:poi}
that it arises very naturally from the braided point of view --
one just has to realise that $x$ is a braided variable. This point
of view also leads to the correct properties of integration.
Namely there is a relevant indefinite integration to go with
$\del_q$ characterised by\cite{KemMa:alg}
\eqn{q-int}{\int_0^{x+y}f(z)\extd_q z=\int_0^y f(z)\extd_q z+\int_0^x
f(z+y)\extd_q z} provided $yx=qxy$, $yz=qzy$ etc., during the
computation. In the limit this gives the infinite Jackson integral
previously known in this context. One also has braided exponentials, braided
Fourier theory etc., for these braided variables.

The braided point of view is also much more powerful than simply
trying to sprinkle $q$ into formulae here and there.

\begin{example} Let $B=\C_q^2$ be the quantum-braided plane generated by
$x,y$ with the relations $yx=qxy$, where two independent copies
have the braid statistics
\[ x'x=q^2xx',\quad x'y=qyx',\quad y'y=q^2yy',\quad
y'x=qxy'+(q^2-1)yx'.\] Here $x',y'$ are the generators of the
second copy of the plane. Then \[ (y+y')(x+x')=q(x+x')(y+y')\]
i.e. $x+x',y+y'$ is another copy of the quantum-braided plane.
Then by similar definitions as above, one has braided partial
derivatives
\[ \del_{q,x}f(x,y)=\frac{f(x,y)-f(qx,y)}{(1-q)x},\quad
\del_{q,y}f(x,y)=\frac{f(qx,y)-f(qx,qy)}{(1-q)y}\]
for expressions normal ordered to $x$ on the left. Note in the
second expression an extra $q$ as $\del_{q,y}$ moves past the $x$
\end{example}

Thus {\em you can add points in the braided plane}, and then (by
an infinitesimal addition) define partial derivatives etc. This is
a problem (multilinear q-analysis) which had been open since 1908
and was only solved relatively recently (by the author) in
\cite{Ma:poi}, as a demonstration of braided mathematics. We note
in passing that $yx=qxy$ is sometimes called the `Manin plane'.
Manin considered only the algebra and a quantum group
action on it, without the braided point of view, without the
braided addition law and without the partial derivatives.

Finally, there is a more formal way by which all such
constructions are done systematically, which we now explain.
It amounts to nothing less than
a new kind of algebra in which algebraic symbols are replaced by
braids and knots.

First of all, given two algebras $B,C$ in a braided
category (such as the representation of $U_q(\cg)$) we have a
braided tensor product $B\und\tens C$ algebra in the same category
defined like a superalgebra but with $-1$ replaced by the braiding
$\Psi_{C,B}$. Thus the tensor product becomes noncommutative (even
if each algebra $B,C$ was commutative) -- the two subalgebras
`commute' up to $\Psi$. This is the mathematical definition of
{\em braid statistics: the noncommutavity of the notion of
`independent' systems}. We call such noncommutativity {\em outer} in
contrast to
the inner noncommutativity of quantisation, which is a property of one
algebra alone. In Example~4, the joint algebra of the independent $x,y$ is
 $\C[x]\und\tens\C[y]$ with
$\Psi(x\tens y)=qy\tens x$. In Example~5 the braided tensor
product is between one copy $x,y$ and the other $x',y'$. The
braiding $\Psi$ in this case is more complicated. In fact it is the same
braiding from the $U_q(su_2)$ spin $\h$ representation that gave the
Jones polynomial. The miracle that makes knot invariants is the
same miracle that allows braided multilinear algebra.

The addition law in both the above examples makes them into
braided groups\cite{Ma:bg}. They are like quantum groups or
super-quantum groups but with braid statistics. Thus, there is a
coproduct
\[ \Delta x=x\tens 1+1\tens x,\quad \Delta y=y\tens 1+1\tens y\]
etc., (this is a more formal way to write $x+x',y+y'$). But
$\Delta:B\to B\und\tens B$ rather than mapping to the usual tensor
product. We do not want to go into the whole theory of braided
groups here. Suffice it to say that the theory can be developed to
the same level as quantum groups: integrals, Fourier theory, etc.,
but using new techniques. One draws the product $B\tens B\to B$ as
a map $\epsfbox{prodfrag.eps}$, the coproduct as
$\epsfbox{deltafrag.eps}$, etc. Similarly with other maps, some
strands coming in for the inputs and some leaving for the outputs.
We then `wire up' an algebraic expression by wiring outputs of one
operation into the inputs of others. When wires have to cross
under or over we have to chose one or the other as $\Psi$ or
$\Psi^{-1}$. We draw such diagrams flowing down the page. An
example of a braided-algebra calculation is given in Figure~5(c).

Braided groups exist in abundance. There are general arguments
that every algebraic quantum field theory contains at its heart
some kind of (slightly generalised) braided group\cite{Ma:int}.
Moreover, the ideas here are clearly very general: braided
algebra.

\subsection{Systematic $q$-Special Relativity}

Clearly braided groups are the correct foundation for q-deformed
geometry based on q-planes and similar q-spaces. One of their main
successes in the period 1992-1994 was a more or less complete and
systematic q-deformation by the team in Cambridge of the main
structures of special relativity and electromagnetism, i.e.
q-Minkowski space and basic structures
\cite{Ma:exa}\cite{MaMey:bra}\cite{Ma:poi}\cite{Ma:fre}\cite{Ma:eps}\cite{KemMa:alg}\cite{Ma:star}\cite{Ma:qsta}\cite{Ma:euc}:

\begin{itemize}
\item q-Minkowski space as $2\times 2$ braided Hermitian matrices

\item q-addition etc., on q-Minkowski space

\item q-Lorentz quantum group $\C_q(SU_2)\dcross \C_q(SU_2)$

\item q-Poincar\'e+scale quantum group $\R_q^{1,3}\lbiprod \widetilde{U_q(so_{1,3})}$

\item q-partial derivatives

\item q-differential forms

\item q-epsilon tensor

\item q-metric

\item q-integration with Gaussian weight

\item q-Fourier theory

\item q-Green functions (but no closed form)

\item q-$*$ structures and q-Wick rotation

\end{itemize}

The general theory works for any braiding or `R-matrix'. I do want
to stress, however, that this project was not in a vacuum. For
example, the algebra of q-Minkowski had been proposed
independently of \cite{Ma:exa} in \cite{CWSSW:ten}, but without
the braided matrix or additive structures. The q-Lorentz was
studied by the same authors but without its quasitriangular
structure, Wess, Zumino et al.\cite{OSWZ:def} studied the
q-Poincar\'e but without its semidirect structure and action on
q-Minkowski space, while Fiore\cite{Fio:sym} studied q-Gaussians
in the Euclidean case, etc. More recently, we
have\cite{Ma:conf}\cite{Ma:dif},

\begin{itemize}
\item q-conformal group $\R_q^{1,3}\lbiprod \widetilde{U_q(so_{1,3})}\rbiprod\R_q^{1,3}$

\item q-diffeomorphism group
\end{itemize}

{\em Notably not on the list, in my opinion still open, is the
correct formulation of the q-Dirac equation.} Aside from this, the
programme came to an end when certain deep problems emerged. In my
opinion they are as follows. First of all, we ended up with formal
power-series e.g. the q-Green function is the inverse Fourier
transform of $(\vec p\cdot \vec p-m^2)^{-1}$ so in principle it is now
defined. But not in closed form! The methods of q-analysis
as in \cite{Jac:fun}\cite{Koo:ort} are not yet far enough advanced to have nice names
and properties for the kinds of powerseries functions encountered. This is a
matter of time. Similarly, braided integration means we can in
principle write down and compute braided Feynman diagrams and
hence define braided quantum field theory at least operatively.
Recently R. Oeckl was able\cite{Oec:bra} to apply the braided
integration theory of \cite{KemMa:alg} not to q-spacetime but directly to a
q-coordinate algebra as the underlying vector space of fields on spacetime.
Here the braided algebra $B$ replaces the
`fields' on spacetime. Choosing a basis of such fields
one can still apply braided
Gaussian integration and actually compute correlation functions.
So the computational problems can and are being overcome.

Secondly and more conceptual, it should be clear that when we
deform classical constructions to braided ones we have to choose
$\Psi$ or $\Psi^{-1}$ whenever wires cross. Sometimes neither will
do, things get tangled up. But if we succeed it means that for
every q-deformation there is another where we could have made the
opposite choice in every case. {\em This classical geometry
bifurcates into two q-deformed geometries according to $\Psi$ or
$\Psi^{-1}$.} Moreover, the role of the $*$ operation is that it
interchanges these two\cite{Ma:qsta}. Roughly speaking,
\[ \begin{matrix}
& \nearrow& q-{\rm geometry}\\ {\rm classical\ geometry}&
&\updownarrow *\\ & \searrow& {\rm conjugate}\ q-{\rm geometry}
\end{matrix}  \]
where the conjugate is constructed by interchanging the braiding
with the inverse braiding (i.e. reversing braid crossings in the
diagrammatic construction). For the simplest cases like the
braided line it means interchanging $q,q^{-1}$. This is rather
interesting given that the $*$ is a central foundation of quantum
mechanics and our concepts of probability. But it also means one
cannot do q-quantum mechanics etc., with q-geometry alone; one
needs also the conjugate geometry.

\subsection{The physical meaning of $q$}

According to what we have said above, the true meaning of $q$ is
that it generalises the $-1$ of fermionic statistics. That is why
it is dimensionless. It is nothing other than a parameter in a
mathematical structure (the braiding) in a generalisation of our
usual concepts of algebra and geometry, going a step beyond
supergeometry.

This also means that q is an ideal parameter for regularising
quantum field theory. Since most constructions in physics
q-deform, such a regularisation scheme is much less brutal than
say dimensional or Pauli-Villars regularisation as it preserves
symmetries as q-symmetries, the q-epsilon tensor etc. \cite{Ma:reg}.
In this context it seems at first {\em too good} a regularisation.
Something has to go wrong for anomalies to appear.
\begin{conjecture}
In q-regularisation the fact that only the Poincar\'e+scale
q-deforms (the two get mixed up) typically results (when the
regulator is removed after renormalisation) in a scale anomaly of
some kind.
\end{conjecture}
This is probably linked to a much nicer treatment of the
renormalisation group that should be possible in this context.
Again a lot of this must
await more development of the tools of q-analysis. At any rate
the result in \cite{Ma:reg} is that q-deformation does indeed
regularise, turning some of the infinities from a Feynman loop integration
into poles $(q-1)^{-1}$.

All of this is related to the Planck scale as follows. Thus, as
well as being a good regulator one can envisage (in view of our
general ideas about noncommutativity and the Planck scale) that
the actual world is in reality better described by $q\ne 1$ due to
Planck scale effects. In other words q-deformed geometry could
indeed be the next-to-classical order approximation to the
geometry coming out of some unknown theory of quantum gravity.
This was the authors own personal reason\cite{Ma:reg} for spending some
years q-deforming the basic structures of physics. The UV cut-off provided by
a `foam-like structure of space time' would instead be provided by
q-regularisation with $q\ne 1$. Moreover, if this is so then
q-deformed quantum field theory should also appear coming out of
quantum gravity as an approximation one better than the usual.
Such a theory would be massless according to the above remarks
(because there is no q-Poincar\'e without the scale generator). Or
at least particle masses would be small compared to the Planck
mass. {\em  How the q-scale invariance breaks would then be a mechanism
for mass generation.}

There are also several other `purely quantum' features of
q-geometry not visible in classical geometry, which would likewise
have consequences for Planck scale physics. One of them is:
\begin{theorem}
 The braided group version of the enveloping algebras $U_q(\cg)$ and
 their $q$-coordinate algebras are isomorphic. I.e. there is
 essentially only one object in q-geometry with different scaling
limits  as $q\to 1$ to give the classical enveloping algebra of $\cg$ or
coordinate algebra of $G$.
\end{theorem}
The self-duality isomorphisms involve dividing by $q-1$ and are
therefore singular when $q=1$, i.e. this is totally alien to
conventional geometric ideas. Enveloping algebras and their
coordinate algebras are supposed to be dual not isomorphic. This
self-duality in q-geometry is rather surprising but is fully
consistent with the self-duality ideas of Section~2. In many ways
q-geometry is simpler and more regular than the peculiar $q=1$
that we are more familiar with.

Recently, it was argued\cite{MajSmo:def} that since loop gravity
is linked to the Wess-Zumino-Witten model, which is linked to
$U_q(su_2)$ (or some other quantum group), that indeed q-geometry
should appear coming out of quantum-gravity with cosomological
constant $\Lambda$. There is even provided a formula
\[ q=e^\frac{2\pi\imath}{2+k},\quad k=\frac{6\pi}{G_{\rm Newton}^2\Lambda}.\]
If so then the many tools of q-deformation developed in the last
several years would suddenly be applicable to study the
next-to-classical structure of quantum-gravity. The fact that loop
variable and spin-network methods `tap into' the revolutions that
have taken place in the last decade around quantum groups, knot
theory and the WZW model (this was evident for example in the
black-hole entropy computation\cite{ABCK:geo}) makes such a conjecture 
reasonable. It also
indicates to me that these new quantum gravity methods are not
just `pushing some problem off to another corner' but are building
on a certain genuine advance that has already revolutionised
several other branches of mathematics. Usually in science when one
big door is opened it has nontrivial repercussions in several
fields.

One way or another the general idea is that quantum effects
dominant at the Planck scale force geometry itself to be modified
as we approach it such as to have a noncommutative
or `quantum' aspect expressed by $q\ne 1$. Although $q$ is
dimensionless and might be given, for example, by formulae such as
the above, one can and should still think of $q$ as behaving
formally like the exponential of an {\em effective Planck's
constant} $\hbar_0$, say. That is we can make semiclassical
expansions, speak of Poisson-brackets being `quantized' etc. This
is not exactly physical quantisation except in so far as quantum
effects at the Planck scale are at the root of it. The precise
physical link can only be made in a full theory of quantum
gravity. It is only in this sense, however, that q-geometry is
`quantum geometry' and `quantum groups' are so called. For
example, the q-coordinate algebras of $U_q(\cg)$ are quantisations
in this sense of a certain Poisson-Lie bracket on $G$ (as
mentioned in Section~4.3). Similarly for all our other q-spaces.

\begin{example}\cite{Ma:conf}
q-Minkowski space quantises a Poisson-bracket on $\R^{1,3}$ given
by the action of the special conformal translations.
\end{example}
This again points to a remarkable interplay between
q-regularisation, the renormalisation group, gravity and particle
mass.

At least in this context we want to note that the braided approach
of this subsection gives a new and systematic approach to the
`quantisation' problem that solves by new `braid diagram' methods
some age-old problems. Usually, one writes a Poisson bracket and
tries to `quantise' it by a noncommutative algebra. Apart from
existence, {\em the problem often overlooked is what I call the
uniformity of quantisation problem}. There is only one universe.
How do we know when we have quantised this or that space
separately that they are consistent with each other, i.e. that
they all fit together to a single quantum universe?

Our theme in Section~2 is of course is that quantisation is not a
well-defined problem. Rather one {\em should} have a deeper point
of view which leads directly to the quantum-algebraic world
-- what we call geometry is then the semiclassical limit of the
intrinsic structure of that, i.e. all different spaces and choices
of Poisson structures on them will emerge from
semiclassicalisation and not vice versa.

Braided algebra solves the uniformity problem in this way. Apart
from giving the q-deformation of most structures in physics, it
does it uniformly and in a generally consistent way because what
what we deform is actually the category of vector spaces into a
braided category. All constructions based on linear maps then
deform coherently and consistently with each other as braid
diagram constructions (so long as they do not get tangled). After
that one inserts the formulae for specific braidings (e.g.
generated by specific quantum groups) to get the q-deformation
formulae. {\em After that} one semiclassicalises by taking
commutators to lowest order, to get the Poisson-bracket that we
have just quantised. Moreover, different quantum groups $U_q(\cg)$
are all mutually consistent being related to each other by an
inductive construction\cite{Ma:dbos}. We have seen this with
q-Minkowski space above.

In summary, the q-deformed examples demonstrate a remarkable
unification of three different points of view; q as a
generalisation of fermionic -1, q as a `quantisation' (so these
ideas are unified) and q as a powerful regularisation parameter in
physics. By the way, these are all far from the original physical
role of q, where $U_q(su_2)$ arose as a generalised symmetry of
the XXY lattice model and where $q$ measures the anisotropy due to
an applied external magnetic field (rather, they are the authors'
point of view developed under the heading of the braided approach
to q-deformation and braided geometry).

\section{Noncommutative differential geometry and Riemannian manifolds}

We have promised that today there is a more or less complete
theory of noncommutative differential geometry that includes most
of the naturally occurring examples such as those in previous
sections, but is a general theory not limited to special examples
and models, i.e. has the same degree of `flabbiness' as
conventional geometry. Here I will try to convince you of this and
give a working definition of a `quantum manifold' and `quantum
Riemannian manifold'\cite{Ma:rie}. I do not want to say that this is
the last word; the subject is still evolving but there is now
something on the table.  Among other things, our constructions are
purely algebraic with operator and $C^*$-algebra considerations as
in Connes' approach not fully worked. In any case, the reader
may well want to start with the more accessible Section~6.4, where
we explore the
semiclassical implications at the more familiar level of the
ordinary differential geometry coming out of the full
noncommutative theory.

\subsection{Quantum differential forms}
As explained in Section~2 our task is
nothing other than to give a formulation of geometry where the
coordinate algebra on a manifold is replaced by a general algebra
$M$. The first step is to {\em choose} the cotangent space or
differential structure. Since one can multiply forms by
`functions' from the left and right, the natural definition is to
define a first order calculus as a bimodule $\Omega^1$ of the
algebra $M$, along with a linear map $\extd:M\to
\Omega^1$ such that
\[ \extd(ab)=(\extd a)b+a\extd b,\quad \forall a,b\in M.\]
Differential structures are not unique even classically, and even
more non-unique in the quantum case. There is, however, one
universal example of which others are quotients. Here
\[ \Omega^1_{\rm univ}=\ker\cdot\subset M\tens M,\quad \extd
a=a\tens 1-1\tens a.\]

Classically we do not think about this much because on a group
there {\em is} a unique translation-invariant differential
calculus; since we generally work with manifolds built on or
closely related to groups we tend to take the inherited
differential structure without thinking. In the quantum case, i.e.
when $M$ is a quantum (or braided) group one has a similar notion
\cite{Wor:dif}: a differential calculus is bicovariant if there
are coactions $\Omega^1\to \Omega^1\tens M, \Omega^1\to M\tens
\Omega^1$ forming a bicomodule and compatible with the bimodule
structures and $\extd$.

\begin{theorem}\cite{Ma:cla}
For the  q-coordinate rings of the quantum groups $U_q(\cg)$,
the (co)irreducible bicovariant $(\omega^1,\extd)$ are essentially
(for generic $q$) in correspondence with the irreducible
representations $\rho$ of $\cg$, and
\[ \Omega^1_{\rm univ}=\oplus_\rho\Omega^1_\rho.\]
\end{theorem}

The lowest spin $\h$ representation of $U_q(su_2)$ defines its
usual differential calculus plus a Casimir as $q\to 1$. The higher
differential calculi show up in the q-geometry and correspond to
higher spin. This should therefore be a step towards understanding
how macroscopic differential geometry arises out of the loop
gravity and spin network formalism. For example, the black-hole
entropy computation\cite{ABCK:geo} reported in Abbay Ashtekar's 
lectures at the conference
was dominated by the spin $\h$ states, which seems to me should be
analogous to the standard differential calculus on the spin
connection bundle dominating as macroscopic geometry emerges from
the quantum gravity theory.

We do not have room to give more details here even of an example
of Theorem~3, but see \cite{Ma:cla}. Instead we content ourselves
with an even simpler and more pedagogical result.

\begin{proposition}\cite{Ma:fie}
If $k$ is a field and $M=k[x]$ the polynomials in one variable,
the (co)irreducible bicovariant calculi $(\Omega^1,\extd)$ are in
correspondence with field extensions of the form
$k_\lambda=k[\lambda]$ modulo $m(\lambda)=0$, where $m$ is an
irreducible monic polynomial. Here
\[ \Omega^1=k_\lambda[x],\quad \extd
f(x)=\frac{f(x+\lambda)-f(x)}{\lambda},\]
\[ f(x).g(\lambda,x)=f(x+\lambda)g(\lambda,x),\quad
g(\lambda,x).f(x)=g(\lambda,x)f(x)\] for functions $f$ and
one-forms $g$.
\end{proposition}

For example, over $\C$, $(\Omega^1,\extd)$ on $\C[x]$ are
classified by $\lambda_0\in C$ and one has
\eqn{diffR}{ \Omega^1=\extd x\C[x],\quad \extd f=\extd
x\frac{f(x+\lambda_0)-f(x)}{\lambda_0},\quad x\extd x=(\extd
x)x+\lambda_0.} We see that the Newtonian case $\lambda_0=0$ is
only one special point in the moduli space of quantum differential
calculi. But if Newton had not supposed that differentials and
forms commute he would have had no need to take this limit. What
one finds with noncommutative geometry is that there is no need to
take this limit at all. In particular, noncommutative geometry
extends our usual concepts of geometry to lattice theory without
taking the limit of the lattice spacing going to zero.

It is also interesting that the most important field extension in physics,
$\R\subset\C$, can be viewed noncommutative-geometrically
with complex functions $\C[x]$ the
quantum 1-forms on the algebra of real functions $\R[x]$. As such its
quantum cohomology is nontrivial, see \cite{Ma:fie}.

\subsection{Bundles and connections}

To go further one has to have a pretty abstract view of
differential geometry. For trivial bundles it is a little easier:
fix a quantum group coordinate ring $H$. Then a gauge field is a
map $H\to \Omega^1$, etc. See \cite{Ma:gos}\cite{BrzMa:dif}. To
define a manifold, however, one has to handle nontrivial bundles.
In noncommutative geometry there is (as yet) no proper way to
build this by patching trivial bundles. All those usual concepts
involve open sets etc, not existing in the noncommutative case.
Fortunately, if one thinks about it abstractly enough one can come
up with a purely algebraic formulation independent of any patches
or coordinate system. For simplicity we are going to limit
attention to the universal calculi; the theory is know for general
calculi as well.

Basically, a classical bundle has a free action of a group and a
local triviality property. In our algebraic terms this
translates\cite{BrzMa:gau}\cite{BrzMa:dif} to an algebra $P$ in
the role of `coordinate algebra of the total space of the bundle',
a coaction $\Delta_R:P\to P\tens H$ of the quantum group $H$ such
that the fixed subalgebra is $M$,
\eqn{fixed}{ M=P^H=\{p\in P|\ \Delta_R p=p\tens 1\}.}
Local triviality is replaced by the requirement that
\eqn{exactness}{ 0\to P(\Omega^1M)P\to\Omega^1P{\buildrel\tilde\chi\over\longrightarrow} P\tens \ker\eps\to 0}
is exact, where $\tilde\chi=(\cdot\tens\id)\Delta_R$ plays the
role of generator of the vertical vector fields corresponding
classically to the action of the group (for each element of $H^*$
it maps $\Omega^1P\to P$ like a vector field). Exactness says that
the one-forms $P(\Omega^1M)P$ lifted from the base are exactly the
ones annihilated by the vertical vector fields.

An example is the quantum sphere. Classically the inclusion
$U(1)\subset SU_2$ in the diagonal has coset space $S^2$ and
defines the $U(1)$ bundle over the sphere on which the monopole
lives. The same idea works here, but since we deal with coordinate
algebras the arrows are reversed. The coordinate algebra of $U(1)$
is the polynomials $\C[g,g^{-1}]$.
\begin{example}
There is a projection from $\C_q(SU_2)\to \C[g,g^{-1}]$
\[\pi\left(\begin{matrix}
a&b\\ c&d
\end{matrix}\right)=\left(\begin{matrix}
g&0\\ 0&g^{-1}
\end{matrix}\right)\]
Its induced coaction $\Delta_R=(\id\tens\pi)\Delta$ is by the
degree defined as the number of $a,c$ minus the number of $b,d$ in
an expression. The quantum sphere $S_q^2$ is the fixed subalgebra
i.e. the degree zero part. Explicitly, it is generated by $b_3=ad$,
$b_+=cd$, $b_-=ab$ with $q$-commutativity relations
\[ b_\pm b_3=q^{\pm 2}b_3b_\pm+(1-q^{\pm 2})b_\pm,
\quad q^{2}b_-b_+=q^{-2}b_+b_-+(q-q^{-1})(b_3-1)\]
and the sphere equation $b_3^3=b_3+qb_-b_+$, and forms a quantum
bundle\cite{BrzMa:gau}\cite{BrzMa:dif}.
\end{example}

When $q\to 1$ we can write $b_\pm=\pm(x\pm\imath y)$, $b_3=z+\h$
and the sphere equation becomes $x^2+y^2+z^2=\frac{1}{4}$ while
the others become that $x,y,z$ commute. The quantum sphere itself
is a member of a 2-parameter family\cite{Pod:sph} of quantum
spheres (the others can also be viewed as bundles in a suitable
framework\cite{BrzMa:geo}.)

One can go on and define a connection as an equivariant splitting
\eqn{connection}{ \Omega^1P=P(\Omega^1 M)P\oplus {\rm complement}}
i.e. an equivariant projection $\Pi$ on $\Omega^1P$. One can show
the required analogue of the usual theory, i.e. that such a
projection corresponds to a connection form such that
\eqn{conform}{\omega:\ker\eps\to\Omega^1P,\quad \tilde\chi\omega=\id}
where $\omega$ intertwines with the adjoint coaction of $H$ on
itself. There is such a connection on the example above -- the
q-monopole\cite{BrzMa:gau}. It is $\omega(g-1)=d\extd a-qb\extd
c$.

Finally, one can define associated bundles. If $V$ is a vector
space on which $H$ coacts then we define the associated `bundles'
$E^*=(P\tens V)^H$ and $E=\hom_H(V,P)$, the space of intertwiners.
The two bundles should be viewed geometrically as `sections' in
classical geometry of bundles associated to $V$ and $V^*$. Given a
suitable (strong) connection one has a covariant derivative
\eqn{covderiv}{ D_\omega:E\to E\tens M,\quad D_\omega=(\id-\Pi)\circ \extd}

This is where the noncommutative differential geometry coming out
of quantum groups links up with the more traditional $C^*$-algebra
approach of A. Connes and others. Traditionally a vector bundle
over any algebra is defined as a finitely generated projective
module. However, there is no notion of quantum principal bundle of
course without quantum groups. The associated bundles to the
q-monopole bundle are indeed finitely generated projective
modules\cite{HajMa:pro}. The projectors are elements of the
noncommutative $K$-theory $K_0(S_q^2)$ and their pairing with
Connes' cyclic cohomology\cite{Con:geo} allows one to show that
the bundle is non-trivial even when $q\ne 1$. Thus the quantum
groups approach is compatible with Connes' approach but provides
more of the (so far algebraic) infrastructure of differential
geometry
-- principal bundles, connection forms, etc. otherwise missing.

\subsection{Soldering and quantum Riemannian structure}

With the above ingredients we can give a working definition of a
quantum manifold. See refer to \cite{Ma:rie} for details. The idea is
that the main feature of being a manifold is that, locally, one can
chose a basis of the tangent space at each point (e.g. a vierbein in physics)
patching up globally via $GL_n$ gauge transformations. In abstract terms it means a frame bundle
 to which the tangent bundle is associated by a `soldering form'.
For a general algebra $M$ we specify this `frame bundle' directly
as some suitable quantum group principal bundle.

Thus, we define a {\em frame resolution} of $M$ as quantum
principal bundle $(P,H,\Delta_R)$ over $M$, a comodule $V$ and an
equivariant `soldering form' $\theta:V\to
P\Omega^1M\subset\Omega^1P$ such that the induced map
\eqn{frame}{ E^*\to \Omega^1M,\quad p\tens v\mapsto p\theta(v)}
is an isomorphism. Of course, all of this has to be done with
suitable choices of differential calculi on $M,P,H$ whereas we
have been focusing for simplicity on the universal calculi. There
are some technical problems here but the same definitions more or
less work in general. Our working definition\cite{Ma:rie} of a
{\em quantum manifold} is this data
$(M,\Omega^1,P,H,\Delta_R,V,\theta)$.

The definition works in that one has many usual results. For example,
a connection $\omega$ on the frame bundle induces a covariant
derivative $D_\omega$ on the associated bundle $E$ which maps over
under the soldering isomorphism to a covariant derivative
\eqn{nabla}{ \nabla:\Omega^1M\to \Omega^1M\tens_M\Omega^1M.}
Its torsion is defined as corresponding similarly to
$D_\omega\theta$.

Defining a Riemannian structure is harder. It turns out that it
can be done in a `self-dual' manner as follows. Given a framing, a
`generalised metric' isomorphism $\Omega^{-1}M\to\Omega^1M$
between vector fields and one forms can be viewed as the existence
of {\em another} framing $\theta^*:V^*\to(\Omega^1M)P$, which we
call the {\em coframing}, this time with $V^*$. Nondegeneracy of
the metric corresponds to $\theta^*$ inducing an isomorphism
$E\isom\Omega^1M$.

Thus our working definition\cite{Ma:rie} of a quantum Riemannian
manifold is the data $(M,\Omega^1,
P,H,\Delta_R,V,\theta,\theta^*)$, where we have a framing and at
the same time $(M,\Omega^1,P,H,\Delta_R,V^*,\theta^*)$ is another
framing. The associated quantum metric is
\eqn{quametric}{ g=\sum_a \theta^*(f^a)\theta(e_a)\in\Omega^1M\tens_M\Omega^1M}
where $\{e_a\}$ is a basis of $V$ and $\{f^a\}$ is a dual basis
(c.f. our friend the canonical element $\exp$ from Fourier theory
in Section~3).

Now, this self-dual formulation of `metric' as framing and
coframing is symmetric between the two. One could regard the
coframing as the framing and vice versa. From our original point
of view its torsion tensor corresponding to $D_\omega\theta^*$ is
some other tensor, which we call the {\em cotorsion
tensor}\cite{Ma:rie}. We then define a generalised Levi-Civita
connection on a quantum Riemannian manifold as the $\nabla$ of a
connection $\omega$ such that the torsion and cotorsion tensors
both vanish.

This is about as far as this programme has reached at present. One
defines curvature of course as corresponding to the curvature of
$\omega$, which is $\extd\omega+\omega\wedge\omega$, but before we
can finish the program outlined in Section~2 we still need to
understand the Ricci and Einstein tensors in this setting. For
this one has to understand their classical meaning more abstractly
i.e. beyond some contraction formulae even in conventional
geometry. It would appear that it has a lot to do with entropy and
the relation between gravity and counting (geometric) states
thermodynamically.

\subsection{Semiclassical limit}

To get the physical meaning of the cotorsion tensor and other
ideas coming out of noncommutative Riemannian geometry, let us
consider the semiclassical limit. {\em What we find is that
noncommutative geometry forces us to slightly generalise
conventional Riemannian geometry itself}. If noncommutative
geometry is closer to what comes out of quantum gravity then this
generalisation of conventional Riemannian geometry should be
needed to include Planck scale effects or at least to be
consistent with them when they emerge at the next order of
approximation.

The generalisation, more or less forced by the noncommutativity,
is as follows:

\begin{itemize}
\item We have to allow any group $G$ in the `frame bundle', hence
the more general concept of a `frame resolution'
$(P,G,V,\theta_\mu^a)$ or {\em generalised manifold}.

\item The {\em generalised metric} $g_{\mu\nu}=\sum_a \theta^*_\mu{}^a\theta_{\nu a}$
corresponding to a coframing $\theta^*_{\mu}{}^a$ is nondegenerate
but need not be symmetric.

\item The {\em generalised Levi-Civita} connection defined as having vanishing torsion and
vanishing cotorsion respects the metric only in a skew sense
\eqn{genlev}{ \nabla_\mu g_{\nu\rho}-\nabla_\nu g_{\mu\rho}=0}

\item The group $G$ is not unique (different flavours of frames are
possible, e.g. an $E_6$-resolved manifold), not necessarily based
on $SO_n$. This gives different flavours of covariant derivative
$\nabla$ that can be induced by a connection form $\omega$.

\item Even when $G$ is fixed and $g_{\mu\nu}$ is fixed, the
generalised Levi-Cevita condition does not fix $\nabla$ uniquely,
i.e. one should use a first order $(g_{\mu\nu},\nabla)$ formalism.
\end{itemize}

To explain (\ref{genlev}) we should note the general result
\cite{Ma:rie} that for any generalised metric one has
\eqn{torcotor}{\nabla_\mu g_{\nu\rho}-\nabla_\nu g_{\mu\rho}={\rm
CoTorsion}_{\mu\nu\rho}-{\rm Torsion}_{\mu\nu\rho},}
where we use
the metric to lower all indices. Here $\omega$ gives two covariant
derivatives
\[ \begin{matrix}
& \theta\nearrow& \nabla\\
\omega& & \\ & \theta^*\searrow& {}^*\nabla
\end{matrix}  \]
depending on whether we regard $\theta$ or $\theta^*$ as the
soldering form. The two are related by
\eqn{nablacompat}{ g({}^*\nabla_X Y,Z)+g(Y,\nabla_XZ)=X(g(Y,Z))}
for vector fields $X,Y,Z$. The cotorsion is the torsion of ${}^*\nabla$.

Our generalisation of Riemannian geometry includes for example
symplectic geometry, where the generalised metric is totally
antisymmetric. So symplectic and Riemannian geometry are included
as special cases and unified in our formulation. This is what we
would expect if the theory is to be the semiclassicalisation of a
theory unifying quantum theory and geometry. It is also remarkable
that metrics with antisymmetric part are exactly what are needed
in string theory to establish T-duality, which is entirely
consistent with our duality ideas of Section~2.

\section*{Acknowledgements}

This was a really great conference in Polanica and I'd like
to thank all of the organisers and participants.

%\bibliographystyle{unsrt}
%\bibliography{biblio}

\baselineskip 15pt

\begin{thebibliography}{10}

\bibitem{Ma:pri}
S.~Majid.
\newblock Principle of representation-theoretic self-duality.
\newblock {\em Phys. Essays}, 4(3):395--405, 1991.

\bibitem{Ma:the}
S.~Majid.
\newblock {\em Non-commutative-geometric Groups by a Bicrossproduct
  Construction}.
\newblock PhD thesis, Harvard mathematical physics, 1988.

\bibitem{Ma:pla}
S.~Majid.
\newblock {H}opf algebras for physics at the {P}lanck scale.
\newblock {\em J. Classical and Quantum Gravity}, 5:1587--1606, 1988.

\bibitem{Ma:rie}
S.~Majid.
\newblock Quantum and braided group {R}iemannian geometry.
\newblock {\em J. Geom. Phys.}, 30:113--146, 1999.

\bibitem{BrzMa:gau}
T.~Brzezi\'nski and S.~Majid.
\newblock Quantum group gauge theory on quantum spaces.
\newblock {\em Commun. Math. Phys.}, 157:591--638, 1993.
\newblock Erratum 167:235, 1995.

\bibitem{Ma:book}
S.~Majid.
\newblock {\em Foundations of Quantum Group Theory}.
\newblock Cambridge Univeristy Press, 1995.

\bibitem{Dri}
V.G. Drinfeld.
\newblock Quantum groups.
\newblock In A.~Gleason, editor, {\em Proceedings of the {ICM}}, pages
  798--820, Rhode Island, 1987. AMS.

\bibitem{Jim:dif}
M.~Jimbo.
\newblock A {$q$}-difference analog of {$U(g)$} and the {Y}ang-{B}axter
  equation.
\newblock {\em Lett. Math. Phys.}, 10:63--69, 1985.

\bibitem{Con:geo}
A.~Connes.
\newblock {\em Noncommutative Geometry}.
\newblock Academic Press, 1994.

\bibitem{Ma:introp}
S.~Majid.
\newblock Beyond supersymmetry and quantum symmetry (an introduction to braided
  groups and braided matrices).
\newblock In M-L. Ge and H.J. de~Vega, editors, {\em Quantum Groups, Integrable
  Statistical Models and Knot Theory}, pages 231--282. World Sci., 1993.

\bibitem{MajSmo:def}
S.~Major and L.~Smolin.
\newblock Quantum deformation of quantum gravity.
\newblock {\em Nucl. Phys. B}, 473:267--290, 1996.

\bibitem{Kli:poi}
C.~Klimcik.
\newblock {P}oisson-{L}ie {T}-duality.
\newblock {\em Nucl. Phys. B (Proc. Suppl.)}, 46:116--121, 1996.

\bibitem{BegMa:poi}
E.~Beggs and S.~Majid.
\newblock {P}oisson-lie {T}-duality for quasitriangular {L}ie bialgebras.
\newblock {\em Commun. Math. Phys.} (in press.)

\bibitem{Ma:ista}
S.~Majid.
\newblock Duality principle and braided geometry.
\newblock {\em Springer Lec. Notes in Phys.}, 447:125--144, 1995.

\bibitem{Ma:reg}
S.~Majid.
\newblock On q-regularization.
\newblock {\em Int. J. Mod. Phys. A}, 5(24):4689--4696, 1990.

\bibitem{Ma:phy}
S.~Majid.
\newblock Physics for algebraists: Non-commutative and non-cocommutative {H}opf
  algebras by a bicrossproduct construction.
\newblock {\em J. Algebra}, 130:17--64, 1990.

\bibitem{BraRob:ope}
O.~Bratteli and D.~W. Robinson.
\newblock {\em Operator Algebras and Quantum Statistical Mechanics}, volume~II.
\newblock Springer-Verlag, 1979.

\bibitem{Ma:rep}
S.~Majid.
\newblock Representations, duals and quantum doubles of monoidal categories.
\newblock {\em Suppl. Rend. Circ. Mat. Palermo, Ser. II}, 26:197--206, 1991.

\bibitem{Ma:som}
S.~Majid.
\newblock Some physical applications of category theory.
\newblock {\em Springer Lec. Notes in Phys.}, 375:131--142, 1991.

\bibitem{Ma:aim}
S.~Majid.
\newblock Non-commutative-geometric {${\cal A}/{\cal G}$}: a new approach to
the quantisation of photons.
\newblock {\em Unpublished}, 1986.

\bibitem{Ma:pho}
S.~Majid.
\newblock Noncommutative geometric quantisation of photons and string field
  theory.
\newblock {\em Unpublished Harvard preprint}, 1988.

\bibitem{Ma:fou}
S.~Majid.
\newblock {F}ourier transforms on {$\cal A/\cal G$} and knot invariants.
\newblock {\em J. Math. Phys.}, 32:924--927, 1990.

\bibitem{Ma:csta}
S.~Majid.
\newblock {$\C$}-statistical quantum groups and {W}eyl algebras.
\newblock {\em J. Math. Phys.}, 33:3431--3444, 1992.

\bibitem{Ash:for}
A.~Ashtekar.
\newblock A 3+1 formulation of {E}instein self-duality.
\newblock {\em Contemp. Math.}, 71:39--53, 1988.

\bibitem{RovSmo:kno}
C.~Rovelli and L.~Smolin.
\newblock Knot theory and quantum gravity.
\newblock {\em Phys. Rev. Lett.}, 61:1155, 1988.

\bibitem{Ma:gau}
S.~Majid.
\newblock Gauge covariant vacuum expectations and glueballs.
\newblock {\em Il Nuov. Cim.}, 104A:949--960, 1991.

\bibitem{MaOec:twi}
S.~Majid and R.~Oeckl.
\newblock Twisting of quantum differentials and the {P}lanck scale {H}opf
  algebra.
\newblock {\em Commun. Math. Phys.}
\newblock 205:617--655, 1999.

\bibitem{Ma:mat}
S.~Majid.
\newblock Matched pairs of {L}ie groups associated to solutions of the
  {Y}ang-{B}axter equations.
\newblock {\em Pac. J. Math.}, 141:311--332, 1990.

\bibitem{MaRue:bic}
S.~Majid and H.~Ruegg.
\newblock Bicrossproduct structure of the {$\kappa$}-{P}oincar{\'e} group and
  non-commutative geometry.
\newblock {\em Phys. Lett. B}, 334:348--354, 1994.

\bibitem{LNRT:def}
J.~Lukierski, A.~Nowicki, H.~Ruegg, and V.N. Tolstoy.
\newblock {$q$}-{D}eformation of {P}oincar{\'e} algebra.
\newblock {\em Phys. Lett. B}, 264,271:331,321, 1991.

\bibitem{AmeMa:wav}
G.~Amelino-Camelia and S.~Majid.
\newblock Waves on noncommutative spacetime and gamma-ray bursts.
\newblock {\em Preprint}, 1999.

\bibitem{ALN:def}
G.~Amelino-Camelia, J.~Lukierski, and A.~Nowicki.
\newblock $\kappa$-deformed covariant phase space and quantum gravity
  uncertainty relations.
\newblock {\em Phys. Atom. Nucl.}, 61:1811--1815, 1988.

\bibitem{Ma:hop}
S.~Majid.
\newblock {H}opf-von {N}eumann algebra bicrossproducts, {K}ac algebra
  bicrossproducts, and the classical {Y}ang-{B}axter equations.
\newblock {\em J. Funct. Analysis}, 95:291--319, 1991.

\bibitem{Ma:varen}
S.~Majid.
\newblock Introduction to braided geometry and {q-M}inkowski space.
\newblock In L.~Castellani and J.~Wess, editors, {\em Proceedings of the
  International School `Enrico Fermi' CXXVII}, pages 267--348. IOS Press,
  Amsterdam, 1996.

\bibitem{Jac:fun}
F.H. Jackson.
\newblock On {$q$}-functions and a certain difference operator.
\newblock {\em Trans. Roy. Soc. Edin.}, 46:253--281, 1908.

\bibitem{Ma:poi}
S.~Majid.
\newblock Braided momentum in the {$q$}-{P}oincar{\'e} group.
\newblock {\em J. Math. Phys.}, 34:2045--2058, 1993.

\bibitem{KemMa:alg}
A.~Kempf and S.~Majid.
\newblock Algebraic $q$-integration and {F}ourier theory on quantum and braided
  spaces.
\newblock {\em J. Math. Phys.}, 35:6802--6837, 1994.

\bibitem{Ma:bg}
S.~Majid.
\newblock Braided groups.
\newblock {\em J. Pure and Applied Algebra}, 86:187--221, 1993.

\bibitem{Ma:int}
S.~Majid.
\newblock Quasi-quantum groups as internal symmetries of topological quantum
  field theories.
\newblock {\em Lett. Math. Phys.}, 22:83--90, 1991.

\bibitem{Ma:exa}
S.~Majid.
\newblock Examples of braided groups and braided matrices.
\newblock {\em J. Math. Phys.}, 32:3246--3253, 1991.

\bibitem{MaMey:bra}
S.~Majid and U.~Meyer.
\newblock Braided matrix structure of {$q$}-{M}inkowski space and
  {$q$}-{P}oincar\'e group.
\newblock {\em Z. Phys. C}, 63:357--362, 1994.

\bibitem{Ma:fre}
S.~Majid.
\newblock Free braided differential calculus, braided binomial theorem and the
  braided exponential map.
\newblock {\em J. Math. Phys.}, 34:4843--4856, 1993.

\bibitem{Ma:eps}
S.~Majid.
\newblock {$q$}-epsilon tensor for quantum and braided spaces.
\newblock {\em J. Math. Phys.}, 36:1991--2007, 1995.

\bibitem{Ma:star}
S.~Majid.
\newblock {$*$}-structures on braided spaces.
\newblock {\em J. Math. Phys.}, 36:4436--4449, 1995.

\bibitem{Ma:qsta}
S.~Majid.
\newblock Quasi-{$*$} structure on {$q$}-{P}oincar{\'e} algebras.
\newblock {\em J. Geom. Phys.}, 22:14--58, 1997.

\bibitem{Ma:euc}
S.~Majid.
\newblock {$q$}-{E}uclidean space and quantum {W}ick rotation by twisting.
\newblock {\em J. Math. Phys.}, 35:5025--5034, 1994.

\bibitem{CWSSW:ten}
U.~Carow-Watamura, M.~Schlieker, M.~Scholl, and S.~Watamura.
\newblock Tensor representation of the quantum group {$SL_q(2,\C)$} and quantum
  {M}inkowski space.
\newblock {\em Z. Phys. C}, 48:159, 1990.

\bibitem{OSWZ:def}
O.~Ogievetsky, W.B. Schmidke, J.~Wess, and B.~Zumino.
\newblock {$q$}-{D}eformed {P}oincar{\'e} algebra.
\newblock {\em Commun. Math. Phys.}, 150:495--518, 1992.

\bibitem{Fio:sym}
G.~Fiore.
\newblock The {$SO_q(N,R)$}-symmetric harmonic oscillator on the quantum
  {E}uclidean space {$R^N_q$} and its {H}ilbert space structure.
\newblock {\em Int. J. Mod. Phys. A}, 26:4679--4729, 1993.

\bibitem{Ma:conf}
S.~Majid.
\newblock Braided geometry of the conformal algebra.
\newblock {\em J. Math. Phys.}, 37:6495--6509, 1996.

\bibitem{Ma:dif}
S.~Majid.
\newblock Quantum and braided diffeomorphism groups.
\newblock {\em J. Geom. Phys.}, 28:94--128, 1998.

\bibitem{Koo:ort}
T.H. Koornwinder.
\newblock Orthogonal polynomials in connection with quantum groups.
\newblock In P.~Nevai, editor, {\em Orthogonal Polynomials: Theory and
  Practice}, number 294 in NATO ASI Series C, pages 257--292. Kluwer, 1990.

\bibitem{Oec:bra}
R.~Oeckl.
\newblock Braided quantum field theory.
\newblock {\em Preprint}, 1999.

\bibitem{ABCK:geo}
A.~Ashtekar, J.~Baez, A.~Corichi, and K.~Krasnov.
\newblock Quantum geometry and black hole entropy.
\newblock {\em Phys. Rev. Lett.}, 80:904--907, 1998.

\bibitem{Ma:dbos}
S.~Majid.
\newblock Double bosonisation and the construction of {$U_q(g)$}.
\newblock {\em Math. Proc. Camb. Phil. Soc.}, 125:151--192, 1999.

\bibitem{Wor:dif}
S.L. Woronowicz.
\newblock Differential calculus on compact matrix pseudogroups (quantum
  groups).
\newblock {\em Commun. Math. Phys.}, 122:125--170, 1989.

\bibitem{Ma:cla}
S.~Majid.
\newblock Classification of bicovariant differential calculi.
\newblock {\em J. Geom. Phys.}, 25:119--140, 1998.

\bibitem{Ma:fie}
S.~Majid.
\newblock Quantum geometry of field extensions.
\newblock {\em J. Math. Phys.}, 40:2311--2323, 1999.

\bibitem{Ma:gos}
S.~Majid.
\newblock Advances in quantum and braided geometry.
\newblock In H.-D. Doebner and V.K. Dobrev, editors, {\em Quantum Group
  Symposium at Group XXI}, pages 11--26. Heron Press, Sofia, 1997.

\bibitem{BrzMa:dif}
T.~Brzezi\'nski and S.~Majid.
\newblock Quantum differentials and the {$q$}-monopole revisited.
\newblock {\em Acta Appl. Math.}, 54:185--232, 1998.

\bibitem{Pod:sph}
P.~Podles.
\newblock Quantum spheres.
\newblock {\em Lett. Math. Phys.}, 14:193--202, 1987.

\bibitem{BrzMa:geo}
T.~Brzezi\'nski and S.~Majid.
\newblock Quantum geometry of algebra factorisations and coalgebra bundles.
\newblock {\em Commun. Math. Phys.} (in press.)

\bibitem{HajMa:pro}
P.~Hajac and S.~Majid.
\newblock Projective module description of the {$q$}-monopole.
\newblock {\em Commun. Math. Phys.}
\newblock 206: 246--264, 1999.


\end{thebibliography}

\end{document}